%% file: main.tex
\documentclass[sigconf, screen]{acmart}
\acmConference[Arxiv]{Arxiv}{April 2021}{Online Archive}
\pdfoutput=1

\setcopyright{none}
\renewcommand\footnotetextcopyrightpermission[1]{}
\input{packages}
\input{macros}

\title{Self-Boosted Automated Program Repair\checkLater{ via\\ On-the-Fly Patch Prioritization}}

\author{Samuel Benton} \affiliation{\institution{The University of Texas at Dallas}} \email{Samuel.Benton1@utdallas.edu}
\author{Mengshi Zhang} \affiliation{Facebook} \email{mengshizhang@fb.com}
\author{Xia Li} \affiliation{\institution{Kennesaw State University}} \email{xli37@kennesaw.edu}
\author{Lingming Zhang}\affiliation{\institution{University of Illinois at Urbana-Champaign}} \email{lingming@illinois.edu}

\begin{document}

\begin{abstract}
    Program repair is an integral part of every software system's lifecycle but can be extremely challenging. To date, researchers have proposed various automated program repair (\apr{}) techniques to reduce efforts of manual debugging. However, given a real-world buggy program, a typical \apr{} technique usually generates a large number of patches, each of which needs to be validated against the original test suite which incurs extremely high computation costs. Although existing \apr{} techniques have already leveraged various static and/or dynamic information to find the desired patches faster, they are still rather costly. In a recent work, researchers proposed \textit{unified debugging} to leverage the patch execution information during \apr{} to help boost fault localization; in this way, the application scope of \apr{} techniques can be extended to all possible bugs, e.g., the patch execution information during \apr{} can help with \textit{manual} repair of the bugs that cannot be automatically fixed. Inspired by unified debugging, this work proposes \tech{} (\textbf{Se}lf-Boosted \textbf{A}utomated \textbf{P}rogram \textbf{R}epair), the first technique to leverage the earlier patch execution information during \apr{} to help boost \textit{automated} repair itself on-the-fly. Our basic intuition is that patches similar to earlier high-quality/low-quality patches should be promoted/degraded to speed up the detection of the desired patches. This experimental study on \toolsActual{} state-of-the-art \apr{} systems demonstrates that, overall, \tech{} can substantially reduce the number of patch 
   executions with negligible overhead. Our study also investigates the impact of various configurations on \tech{}. Lastly, our study demonstrates that \tech{} can even leverage the patch execution information from other \apr{} tools from the same buggy program to further boost \apr.
\end{abstract}

\maketitle

\input{paper-sections/01_introduction}
\input{paper-sections/06_related-works}
\input{paper-sections/03_approach}
\input{paper-sections/04_study-design}
\input{paper-sections/05_analysis}

\input{paper-sections/07_conclusion}

\bibliographystyle{ACM-Reference-Format}
\bibliography{bibliographies/mendeley, bibliographies/citations-other}

\end{document}

%% file: packages.tex
\usepackage[utf8]{inputenc}

\usepackage{balance}
\usepackage{caption}
\usepackage{listings}
\usepackage{comment}
\usepackage{color, colortbl}
\usepackage{xcolor}
\usepackage{xspace}
\usepackage{multirow}
\usepackage{mdframed}
\usepackage{amsmath}
\usepackage[linesnumbered,ruled,vlined]{algorithm2e}
\usepackage{subcaption}

\usepackage{ulem}
\usepackage{hhline}

%% file: macros.tex
\definecolor{lightblue}{rgb}{0.25, 0.25, 0.75}
\definecolor{lightred}{rgb}{1.00, 0.75, 0.75}
\definecolor{darkred}{rgb}{0.75, 0.25, 0.25}
\definecolor{tablegray}{gray}{0.95}

\newcommand{\toolsConsidered}{16}
\newcommand{\toolsActual}{12}

\newcommand{\tp}{\textit{e$_{f}$}}
\newcommand{\fp}{\textit{n$_{f}$}}
\newcommand{\tn}{\textit{e$_{p}$}}
\newcommand{\fn}{\textit{n$_{p}$}}
\newcommand{\tftuple}{similarity tuples}

\newcommand{\dr}{patch reduction}

\newcommand{\sam}[1]{{\colorbox{lightblue}{\color{white}Sam:}\color{lightblue}#1}}
\newcommand{\highlight}[1]{{\color{darkred}TO UPDATE: #1}}

\newcommand{\apr}{APR\xspace}
\newcommand{\tech}{SeAPR\xspace}
\newcommand{\techPat}{SeAPR++\xspace}
\newcommand{\profl}{ProFL}
\newcommand{\unidebug}{UniDebug$_{++}$}
\newcommand{\dfts}{Defects4J}

\newcommand{\baselineOld}{P$_{baseline}$}
\newcommand{\baselineNew}{P$_{new}$}

\newcommand{\arja}{Arja}
\newcommand{\kali}{Kali-A}
\newcommand{\genprog}{GenProg-A}
\newcommand{\rsrepair}{RSRepair-A}
\newcommand{\cardumen}{Cardumen}
\newcommand{\jgenprog}{JGenProg}
\newcommand{\jkali}{JKali}
\newcommand{\jmutrepair}{JMutRepair}
\newcommand{\dynamoth}{DynaMoth}
\newcommand{\nopol}{Nopol}
\newcommand{\avatar}{Avatar}
\newcommand{\kpar}{KPar}
\newcommand{\fixminer}{FixMiner}
\newcommand{\tbar}{TBar}

\newcommand{\simfix}{Simfix}
\newcommand{\acs}{Acs}

\newcommand{\Comment}[1]{}
\newcommand{\checkLater}[1]{}
\newcommand{\lingming}[1]{\colorbox{green}{\textcolor{red}{LINGMING:}}\textcolor{red}{#1}}
\newcommand{\mengshi}[1]{\colorbox{blue}{\textcolor{yellow}{Mengshi:}}\textcolor{red}{#1}}
\newcommand{\Xia}[1]{{\color{blue}#1}}

\newtheorem{theorem}{\textbf{Definition}}[section]

\newcounter{finding}
\newenvironment{finding}
    {
        \refstepcounter{finding}
    	\begin{mdframed}[
    	nobreak=true,
    	linecolor=black,
    	roundcorner=12pt,
    	backgroundcolor=tablegray,
    	linewidth=0.5pt,
    	leftmargin=0.5em,
    	rightmargin=0.5em,
    	topline=true,
    	bottomline=true,
    	frametitlerule=true,
    	frametitlebackgroundcolor=gray!30,
    	frametitlerulecolor=gray,
    	frametitle=Finding \arabic{finding},
    	frametitleaboveskip=0.2em,
    	frametitlebelowskip=0.1em,
    	skipabove=12pt
    	]
 	}
 	{
 	\end{mdframed}
    \vspace{1em}
    }

\newcounter{equationz}

\newcommand{\distance}{0.5em}
\newcommand{\mathmp}{$\mathbf{M_p}$}
\newcommand{\mathmm}{$\mathbf{M_m}$}
\newcommand{\mathe}{$\mathbf{e}$}
\newcommand{\mathE}{$\mathbf{E}$}

\newcommand{\mathr}{$\mathbf{r}$}
\newcommand{\mathR}{$\mathbf{R}$}

\newcommand{\mathmv}{$\mathbf{M_v}$}
\newcommand{\matht}{$\mathbf{t}$}
\newcommand{\mathT}{$\mathbf{T}$}
\newcommand{\mathp}{$\mathbf{p}$}
\newcommand{\mathP}{$\mathbf{P}$}

\newcommand{\skipped}{-}
\newcommand{\passed}{\checkmark{}}
\newcommand{\failed}{$\times$}

\newcommand{\mathpb}{$\mathbf{p_{b}}$} 
\newcommand{\mathPh}{$\mathbf{P_{h}}$}
\newcommand{\mathPl}{$\mathbf{P_{l}}$}
\newcommand{\mathCell}[3]{#1[#2,#3]}
\newcommand{\mathRow}[2]{#1[#2]}

\newcommand{\setdiff}{$\ominus$}

\newcommand{\patcha}{\mathp$_1$}
\newcommand{\patchb}{\mathp$_2$}
\newcommand{\patchc}{\mathp$_3$}
\newcommand{\patchd}{\mathp$_4$}

\newcommand{\tsta}{$t_1$}
\newcommand{\tstb}{$t_2$}
\newcommand{\tstc}{$t_3$}

\newcommand{\elea}{$e_1$}
\newcommand{\eleb}{$e_2$}
\newcommand{\elec}{$e_3$}
\newcommand{\eled}{$e_4$}

\newcommand{\pata}{$r_1$}
\newcommand{\patb}{$r_2$}
\newcommand{\patc}{$r_3$}

\newcommand{\res}{$\mathbf{r}$}
\newcommand{\plau}{$\mathbf{P_{\passed{}}}$}

\newcommand{\ochiai}{Ochiai}
\newcommand{\pp}{pp}


%% file: paper-sections/01_introduction.tex
\section{Introduction}
\label{section:introduction}



Software systems persist everywhere in all facets of today's society; they drive financial institutions, facilitate communication worldwide, oversee critical systems, and so forth. Software systems, however, are \textit{frequently distributed} with numerous bugs that will eventually lead to severe disasters. For example, 
the 5th edition of \texttt{Tricentis.com}'s annual report (2017) shows that software failures impact half of the world’s population (3.7 billion people) and \$1.7 trillion in assets; it also mentiones that there can be far more software
bugs in the wild than we will likely ever know about~\cite{Tricentis}. 
Therefore, in practice, it is imperative for developers to fix these bugs as early as possible with minimal resource consumption. However, manual bug fixing can be extremely tedious, challenging, and time-consuming since modern software systems can be extremely complicated~\cite{bib:Cambridge}.

Fortunately, in lieu of manual bug fixing, researchers have also extensively studied Automated Program Repair (\apr{})~\cite{Wang2020AutomatedWe,toolTbar,toolAvatar,toolDynamoth,toolCardumen,Jiang2018ShapingCode,toolAstor, toolArja, Lou2019History-drivenWe, Wu2019AutomatingTransformation,Jiang2019InferringCode, Long2016AutomaticCode, Qi2015AnSystems, Long2015StagedSynthesis, Long2017AutomaticGeneration, Long2016AnSystems}, which aims to automatically fix software bugs to reduce manual debugging efforts. Typical \apr{} techniques leverage off-the-shelf fault localization~\cite{Wong2016ALocalization} techniques (such as Ochiai~\cite{Abreu2007OnLocalization} and Tarantula~\cite{Jones2005EmpiricalTechnique}) to identify potential buggy locations. Then, they leverage various techniques to generate potential software patches for the potential buggy locations. Lastly, each generated patch will need to be executed against the original test suite to find the \textit{plausible} patches (i.e., the patches that can pass all the original tests). Note that not all the plausible patches are the patches that the developers want; thus, developers need to further inspect the produced plausible patches to derive the final \textit{correct} ones. To date, various \apr{} techniques have been proposed, including techniques based on predefined-templates~\cite{toolTbar, toolPrapr, Wen2018Context-awareRepair}, heuristics~\cite{Jiang2018ShapingCode, toolArja, LeGoues2012AEach}, and constraint solving~\cite{toolNopol, Xiong2017PreciseRepair, toolCardumen}. Furthermore, \apr{} techniques have also drawn wide attention from the industry, e.g., Facebook~\cite{Marginean2019SapFix:Scale}, Fujitsu~\cite{Saha2017ELIXIRRepair}, and Alibaba~\cite{techniqueProfl}.


Compared with manual bug fixing, \apr{} can automatically fix a number of real-world bugs within minimal human intervention and can be easily integrated with the natural workflow of continuous integration lifecycle (e.g., Facebook's in-house tool SapFix~\cite{Marginean2019SapFix:Scale} has been integrated into their workflow such that any modification to the underlying system are automatically checked for bugs and fixed if possible). Meanwhile, despite the promising future of \apr{}, 
it is not a perfect science and numerous issues still plague the area. Among the most paramount of these issues are still the time costs associated with attempting numerous patches for large-scale real-world systems. Existing studies have demonstrated that the patch validation costs dominate the \apr{} costs~\cite{Chen2020FastAll, le2013current, weimer2013leveraging, mehne2018accelerating}, since each patch needs to be executed against the original test suite.
 
To reduce the \apr{} costs, researchers have proposed various techniques to reduce the number of patches generated, e.g., based on machine learning~\cite{Saha2017ELIXIRRepair}, code mining~\cite{Jiang2019InferringCode}, and constraint solving~\cite{toolNopol}. However, prior work has demonstrated that such techniques can incur the dataset overfitting issue, i.e., the correct patches may be pruned for many other unstudied cases~\cite{toolPrapr}. Furthermore, researchers have also proposed to prioritize all the generated patches to find the plausible patches earlier. Such existing techniques primarily utilize static or dynamic information to \textit{statically} prioritize patches before the patch validation, e.g., most \apr{} techniques use the suspiciousness values computed by off-the-shelf fault localization techniques to prioritize the patch validation. No further reprioritization is employed during the patch validation process of these tools, leading to limited improvement.

In a recent work, researchers proposed \textit{unified debugging}~\cite{techniqueProfl, Benton2020} to leverage the patch execution information during \apr{} to help boost fault localization; in this way, the application scope of \apr{} techniques can be extended to all possible bugs, e.g., the patch execution information during \apr{} can help with \textit{manual} repair of the bugs that cannot be automatically fixed. Inspired by unified debugging, this work proposes \tech{} (\textbf{Se}lf-Boosted \textbf{A}utomated \textbf{P}rogram \textbf{R}epair), the first technique to leverage the patch-execution information during \apr{} to help boost \textit{automated} repair itself on-the-fly. Our basic intuition is that earlier patch execution results can help better prioritize later patch executions on-the-fly to speed up the detection of the desired patches (e.g., plausible/correct patches). In this way, we promote the ranking of the patches similar to the executed high-quality patches, while degrading the ranking of the patches similar to the executed low-quality patches. More specifically, we analyze the modified elements of patches to compute the similarity among patches as patches modifying similar program elements can exhibit close program behaviors. This experimental study on \toolsActual{} state-of-the-art \apr{} systems demonstrates that, overall, \tech{} can reduce the number of patch validations by up to 78\%. Our study also investigates the impact of various configurations of \tech{}, e.g., the formula for patch prioritization, the type of patch-execution matrices (full or partial), the granularity of modified elements considered for similarity computation (e.g., packages, classes, methods, or statements), and additional patch execution information for computing patch similarity. Lastly, our study demonstrates that \tech{} can even leverage the patch-execution information from other \apr{} tools for the same buggy program to further boost the current \apr{} tools.

To summarize, this paper makes the following contributions:
\begin{itemize}
        \item \textbf{Direction.} This paper opens a new dimension to leverage patch-execution information to boost automated program repair on-the-fly and can inspire more future work in this new direction.
        \item \textbf{Design.} We design the first technique, \tech{}, in this new direction to update each patch's priority scores based on its similarity with the executed patches and the quality of the executed patches.
        \item \textbf{Extensive Study.} We have performed an extensive study of the proposed technique on \toolsActual{} state-of-the-art \apr{} systems for JVM-based languages using the widely studied real-world bugs from Defects4J.
        \item \textbf{Practical Guidelines.} The study reveals various practical guidelines, including (1) \tech{} can substantially speed up the studied \apr{} techniques by up to 78\% with negligible overhead (e.g., less than 0.5 seconds per project), (2) \tech{} has stable performance when using different formulae for computing patch priority and different types of patch-execution matrices, (3) finer-grained modification information and additional fixing pattern information for patch similarity computation can both further boost \tech{}, and (4) \tech{} can effectively utilize historical patch-validation results from other \apr{} tools to boost current \apr{} tools.
    \end{itemize}


\Comment{Section \ref{section:approach} details our underlying approach. Section \ref{section:design} describes our experimental design and evaluation metrics. Finally, section \ref{section:analysis} delves deeper into the detailed analysis of our results. \mengshi{do we need this?}}

%% file: paper-sections/06_related-works.tex
\section{Related Work}
\label{section:related}

\subsection{Automated Program Repair}
Automated Program Repair (\apr{}) techniques~\cite{Wang2020AutomatedWe, toolTbar, toolAvatar, toolDynamoth, toolCardumen, Jiang2018ShapingCode, toolAstor, toolArja, Lou2019History-drivenWe, Wu2019AutomatingTransformation, Jiang2019InferringCode, Long2016AutomaticCode, Qi2015AnSystems, Long2015StagedSynthesis, Long2017AutomaticGeneration, Long2016AnSystems} aim to automatically fix software bugs to substantially reduce manual debugging efforts and have been extensively studied during the last decade. Existing \apr{} techniques can be categorized into two broad families: (1) techniques that monitor the dynamic program executions to find runtime violations of certain specifications and then \textit{heal} the program via directly modifying the program runtime state in case of unexpected program behaviors~\cite{Long2014AutomaticShepherding, Perkins2009AutomaticallySoftware} and (2) techniques that modify program code representations based on various patch-generation techniques and then validate each generated patch (e.g., via testing~\cite{LeGoues2012AEach}, formal specification checking~\cite{Pei2014AutomatedContracts}, and static analysis~\cite{vanTonder2018StaticProperties}) to find the final desired patches. In recent years, the code-representation-level techniques, especially those that leverage testing for patch validation, have gained popularity as testing is the dominant methodology for detecting software bugs in practice. Such typical \apr{} techniques usually include the following phases. (1) \textit{Fault localization}: \apr{} techniques first leverage off-the-shelf fault localization techniques~\cite{toolCardumen,toolArja,toolAvatar,toolFixminer,Liu2019YouSystems,toolTbar} to localize the potential buggy locations. (2) \textit{Patch generation}: \apr{} techniques will leverage various strategies to generate potential patches for the identified potential buggy locations. (3) \textit{Patch validation}: all the generated patches will be executed against the original test suite to detect the patches that can pass all the original tests, i.e., \textit{plausible patches}. Of course, since not all plausible patches are desirable, patch correctness checking (often done via manual inspection in practice) needs to be further performed to find the final \textit{correct patches}, i.e., the patches equivalent to developer patches.

According to a recent study~\cite{Liu2020OnPrograms}, state-of-the-art \apr{} techniques can be mainly divided into the following categories. (1) \textit{Heuristic-based techniques} leverages various heuristics to iteratively explore the search space of all possible program edits. For example, the seminal GenProg technique~\cite{LeGoues2012AEach} leverages genetic programming to synthesize donor code for high-quality patch generation, while the recent SimFix technique~\cite{Jiang2018ShapingCode} employs advanced code search to obtain donor code for patch generation. (2) \textit{Template-based techniques} leverage predefined fixing templates (e.g., changing ``>'' to ``>='') to perform patch generation. Such predefined fixing templates can be either manually summarized (e.g., KPar~\cite{Kim2013AutomaticPatches}), or automatically inferred (e.g., HDRepair~\cite{Le2016HistoryRepair}) from historical bug fixes. (3) \textit{Constraint-based techniques} transform the program repair problem into a constraint-solving problem and leverage state-of-the-art constraint solvers (e.g., SMT~\cite{DeMoura2008Z3:Solver}) for patch generation. For example, Nopol~\cite{toolNopol} leverages SMT solvers to encode the bug-correction constraints and further solves the constraints to synthesize simple expressions to fix buggy conditional statements.
More recently, researchers have also looked into \textit{learning-based techniques}~\cite{Lutellier2020CoCoNuT:Repair, Ding2020PatchingMetaphor} to directly generate code patches via learning from historical fixes.

\subsection{Cost Reduction for \apr{}}
Despite the promising future of \apr{}, it can be extremely time consuming due to the generation and validation of a large number of possible patches. Actually, the patch validation cost has been shown to dominate the overall \apr{} cost~\cite{Chen2020FastAll, Mehne2018AcceleratingRepair}. Therefore, researchers have also looked into various techniques to further speed up \apr{}. To reduce \textit{the validation time spent on each patch}, Ghanbari et al.~\cite{toolPrapr} and Chen et al.~\cite{Chen2020FastAll} proposed to share the same JVM session across multiple patch validations; in this way, the patch loading and execution time can be substantially accelerated for both source-code and bytecode level \apr{} techniques. In addition, researchers have also proposed to prioritize and reduce the test executions for each patch to reduce the validation time for each patch. For example, Qi et al.~\cite{Qi2013EfficientPrioritization} proposed TrpAutoRepair to prioritize test executions for each patch based on history information to falsify implausible patches faster; Mehne et al.~\cite{Mehne2018AcceleratingRepair} further proposed to reduce the number of test executions for each patch, since tests not covering the patched location(s) cannot help validate the patch. To reduce \textit{the number of validated patches}, almost all existing \apr{} techniques leverage fault localization and various other strategies to reduce the possible patch executions. Furthermore, many existing \apr{} techniques also leverage other available dynamic or static information to prioritize patch executions to find the desired patches faster (e.g., based on various fault localization~\cite{Wen2018Context-awareRepair, Abreu2007OnLocalization}).\checkLater{ For example, \sam{all of the \toolsConsidered{}} \lingming{need to add how the 13 studied apr tools prioritize patches}} Despite various cost reduction techniques have been proposed, \apr{} techniques are still rather time consuming for real-world programs~\cite{toolPrapr}. In this paper, we propose the first technique to leverage on-the-fly patch execution information to help better prioritize patch executions. Note that our technique is orthogonal to all existing patch prioritization techniques and our experimental results demonstrate that our technique can substantially speed up state-of-the-art \apr{} techniques with various patch prioritization strategies.

\subsection{Unified Debugging}
In the past decade, fault localization has been widely used to help \apr{} identify potential buggy locations for patching. The recent \textit{unified debugging} work~\cite{techniqueProfl, Benton2020} unifies the traditional process fault localization and \apr{} in the other direction for the first time. The basic intuition of unified debugging is that the massive patch execution information during \apr{} can actually substantially boost fault localization. For example, if a patch can pass all the tests, it means the patch likely mutes the impacts of the bug, even though this patch that may not correct; it can then be inferred that the patched location is highly related to the actual buggy location, since otherwise the bug impact would not be muted. 
With unified debugging, even when \apr{} techniques cannot fix a bug, unified debugging still analyzes the patch-execution information to provide useful hints about potential buggy locations to help with \textbf{\textit{manual repair}}. In this way, unified debugging can extend the application scope of \apr{} to all possible bugs, not only the bugs that are automatically fixable.

Inspired by the unified debugging work, we also aim to leverage the wealth of patch execution information during \apr{}. Meanwhile, there are the following major differences between our work and unified debugging. First, while unified debugging aims to leverage patch execution information for \textit{manual} program repair, our technique aims to leverage such information to directly boost \textit{automated} program repair, i.e., \textbf{we aim to improve the performance of \apr{} tools by prioritizing high-quality patches sooner in the validation process}. Second, technique principles\checkLater{ \highlight{???}} are also substantially different. Unified debugging analyzes \textit{the correlation between patch locations and test outcomes} to infer potentially buggy locations, while our work analyzes \textit{the correlation among executed and remaining patches} via estimating their behavioral similarities based on their modified code elements to speed up the detection of desired patches. In fact, our patch-similarity estimation idea is inspired by an early work on mutation testing~\cite{Zhang2013FasterReduction}, which leverages the similarities of modified elements of each mutant to perform test prioritization and reduction for each mutant to speed up mutation testing.

%% file: paper-sections/03_approach.tex
\section{Studied Approach}
\label{section:approach}

In this section, we first present the necessary preliminaries and basics for designing our \tech{} approach (Section~\ref{section:pre}). Then, we introduce the detailed \tech{} approach (Section~\ref{section:basic}). Furthermore, we will also talk about different variants of \tech{} (Section~\ref{section:variants}). Lastly, we will introduce a further extension of \tech{} to leverage the patch execution information from other \apr{} tools for even faster \apr{} (Section~\ref{section:extensions}).

\subsection{Preliminaries}
\label{section:pre}
Before further detailed discussion of our approach, we now introduce the key definitions and concepts in this section.

\input{paper-sections/definitions/def-validation-matrix}

Ideally the patch-validation matrix should be \textit{full}, i.e., every cell should be \passed{} or \failed{}. In practice, during the \apr{} process, however, most modern \apr{} tools terminate the test execution for one patch immediately after observing any failing test on that particular patch, since the primary goal is to find correct patches and patches which fail any test must not even be plausible. In this way, the \apr{} process can be largely sped up without sacrificing repair effectiveness. Meanwhile, not all \apr{} tools employ this strategy so we study both types of matrices, where some tests remain unexecuted (\textbf{partial matrices}) versus where all tests always execute (\textbf{full matrices}). Table~\ref{figure:quality-matrix} presents the example full and partial matrices for four example patches (i.e., \patcha{}, \patchb{}, \patchc, and \patchd{}) on three example tests (i.e., \tsta, \tstb, and \tstc). Note that the first row for the patch-validation matrix is always the test execution results of the original buggy program (i.e., \mathpb). 

\input{paper-sections/definitions/def-modification-matrix}

Note that the patch modification matrix can be defined at different levels (e.g., at the package, class, method, and statement granularities) depending on the granularity of considered program elements. For example, at the default method granularity, the columns will be the methods modified by each program patch. Table~\ref{figure:similarity-elements} presents an example patch modification matrix for the above four example patches on four program elements (i.e., \elea{}, \eleb{}, \elec{}, and \eled{}). In this way, since \patcha{} patches program elements \{\elea{}, \eleb{}, \elec{}\}, the first three columns are \passed{} for \patcha{}.


\subsection{Basic \tech}
\label{section:basic}
Given the above introduced patch-validation matrix and patch-modification matrix (which are readily available for almost all \apr{} tools) for the already executed/validated patches, our \tech{} performs on-the-fly patch prioritization to speed up \apr. Our basic intuition is that patches similar to executed high-quality patches are likely to also be high-quality and should therefore be prioritized earlier; likewise, patches rather similar to executed low-quality patches should be deprioritized as they are likely to produce low-quality patches. In this section, we first introduce our definitions for patch quality (Section~\ref{section:patch_quality}); then, we introduce the detailed strategy to compute patch similarity with high- or low-quality patches (Section~\ref{section:patch_similarity}); we introduce our final priority score computation for all unexecuted patches (Section~\ref{section:patch_pri}); lastly, we present our overall algorithm (Section~\ref{section:alg}) with corresponding examples (Section~\ref{section:example}).

\subsubsection{Patch Quality}
\label{section:patch_quality}
When processing the patches that have been executed/validated, we need to estimate the patch's quality by analyzing the patch validation matrix. Intuitively, the ultimate goal of \apr{} is to produce plausible patches that can pass all the original tests. Therefore, whether a patch passes some originally failing tests can be used to estimate the patch quality. Therefore, in this study, a patch is classified as high-quality (patches we wish to prioritize) if it can make \textit{any originally failing test pass}; likewise, a patch is classified as low-quality (patches we wish to deprioritize) if it cannot make any originally failing test pass. Formally, the set of high-quality and low-quality patches can be defined as Equation (\ref{eq:ph}) and (\ref{eq:pl}), respectively.

 \begin{equation}
    \text{\mathPh{} = \{ \mathp{} | $\exists$ \matht{}, \mathCell{\mathmv{}}{\mathp{}}{\matht{}} = \passed{} $\wedge{}$ \mathCell{\mathmv{}}{\mathpb{}}{\matht{}} = \failed{} \}}
    \label{eq:ph}
\end{equation}

\begin{equation}
    \text{\mathPl{} = \{ \mathp{} | $\forall$ \matht{}, \mathCell{\mathmv{}}{\mathpb{}}{\matht{}} = \failed{} $\Rightarrow{}$
    \mathCell{\mathmv{}}{\mathp{}}{\matht{}} $\neq{}$ \passed{} \}}
    \label{eq:pl}
\end{equation}


Note that we can also easily compute the detailed number of originally failing tests that now pass on a patch; however, prior work has demonstrated that the detailed test number can be misleading~\cite{techniqueProfl}. Of course, this is just the first work in this new direction, and we highly encourage other researchers to investigate other better ways to estimate patch quality.

\subsubsection{Patch Similarity}
\label{section:patch_similarity}
After calculating patch quality for executed patches, we iterate through all remaining patches within \mathP{} to compute their similarity information with the executed high/low-quality patches. For each patch \mathp{} that has not been validated yet, we compare its patch modification matrix information against that of each of the validated patches. During the comparison, we compute the number of elements \textit{matching} and \textit{differing} among the two compared patches (i.e., two rows in the patch modification matrix). We calculate the number of matching elements by performing the \textit{set intersection} on the two patch modification matrix rows representing the two patches. Likewise, we calculate the number of differing elements by performing a \textit{symmetric set difference} (i.e., A \setdiff{} B = (A - B) $\cup{}$ (B - A)) on the two patch modification matrix rows representing the two patches. It is important to note that we employ the \textit{symmetric} but not \textit{asymmetric} set difference, since this ensures that the similarity between two patches is symmetric.

Based on the similarity/dissimilarity with high/low-quality patches, we can compute the following tuple for each unvalidated patch \mathp{} for prioritization, (\tp{}, \fp{}, \tn{}, \fn{}). Our basic idea is that \tp{} should get increased when \mathp{} shares elements with high-quality patches, \tn{} should get increased when \mathp{} shares elements with low-quality patches, \fp{} should get increased when \mathp{} has set difference with high-quality patches, and \fn{} should get increased when \mathp{} has set difference with low-quality patches. Since the detailed number of the matching/different modified elements between two patches can tell the detailed similarity/dissimilarity information, the increment should also consider such detailed information. In this way, the formulae for computing the tuple for each unvalidated patch \mathp{} are shown in the following equations: 

\begin{equation}
        \label{equation:tp}
         \text{\tp{}[\mathp{}] = $\sum{}_{\mathbf{p}'}${ | \{ \mathe{} | \mathe{} $\in$ \mathRow{\mathmm{}}{\mathp{}} $\cap$ \mathRow{\mathmm{}}{\mathp{}'} $\wedge$ \mathp{}' $\in$ \mathPh{} \}|}} 
\end{equation}

\begin{equation}
        \label{equation:tn}
         \text{\tn{}[\mathp{}] = $\sum{}_{\mathbf{p}'}${ | \{ \mathe{} | \mathe{} $\in$ \mathRow{\mathmm{}}{\mathp{}} $\cap$ \mathRow{\mathmm{}}{\mathp{}'} $\wedge$ \mathp{}' $\in$ \mathPl{} \}|}} 
\end{equation}

\begin{equation}
        \label{equation:fp}
         \text{\fp{}[\mathp{}] = $\sum{}_{\mathbf{p}'}${ | \{ \mathe{} | \mathe{} $\in$ \mathRow{\mathmm{}}{\mathp{}} \setdiff{} \mathRow{\mathmm{}}{\mathp{}'} $\wedge$ \mathp{}' $\in$ \mathPh{} \}|}} 
\end{equation}

\begin{equation}
        \label{equation:fn}
         \text{\fn{}[\mathp{}] = $\sum{}_{\mathbf{p}'}$ { |\{ \mathe{} | \mathe{} $\in$ \mathRow{\mathmm{}}{\mathp{}} \setdiff{} \mathRow{\mathmm{}}{\mathp{}'} $\wedge$ \mathp{}' $\in$ \mathPl{} \}|}} 
\end{equation}

Note that \mathRow{\mathmm}{\mathp} denotes the set of program elements modified by patch \mathp. For example, if a validated patch \mathp' is high-quality and shares elements with the current \mathp, the \tp{} of \mathp{} is then increased for |\mathRow{\mathmm{}}{\mathp{}} $\cap$ \mathRow{\mathmm{}}{\mathp{}'}|. In this way, all the other tuple elements can be defined in a similar way.

\subsubsection{Patch Prioritization}
\label{section:patch_pri}

Based on the similarity tuple we computed from the previous step, we can compute the priority for each unvalidated patch based on the following intuition: (1) a patch more similar/dissimilar with high-quality patches should be promoted/degraded, (2) a patch similar/dissimilar with low-quality patches should be degraded/promoted. Actually, such intuition is quite similar to traditional spectrum-based fault localization~\cite{Jones2002VisualizationLocalization} where the intuition is (1) a program element executed/unexecuted by failed tests should be more/less suspicious, (2) a program elements executed/unexecuted by passed tests should be less/more suspicious. In this way, all the traditional fault localization formulae can be directly leveraged here to compute the patch priority. We use the Ochiai formula, shown in Equation (\ref{formula:Ochiai}), as our default formula as it is  often the default formula for spectrum-based fault localization techniques~\cite{toolPrapr, Wen2018Context-awareRepair, Jiang2018ShapingCode}. In this way, patches will be promoted/demoted if they are similar/dissimilar with other high-quality patches, consistent with our intuition. \checkLater{In this way, a patch\checkLater{ \sam{should vs will?}} will get higher/lower rank if it is similar/dissimilar with high-quality patches, .}

\begin{equation}
   \text{Ochiai} = \dfrac{\tp{}}{\sqrt{(\tp{} + \fp{}) * (\tp{} + \tn{})}}
    \label{formula:Ochiai}
\end{equation}

\subsubsection{Overall Algorithm}
\label{section:alg}
Given the above definitions, we can now present the overall \tech{} algorithm. Shown in Algorithm~\ref{alg:seapr}, \tech{} first initializes the \tftuple{} for all patches as 1s (Line 1). Then, \tech{} iterates through all patches and validates them in order (Lines 3-16). During each iteration, \tech{} first gets the patch \mathp{} with the highest priority and removes that from the patch list \mathP. Note that for the patches with tied \tech{} priority scores \checkLater{\sam{check consistent of "priority score"}} (e.g., all patches are tied before the first patch execution), \tech{} prioritizes them with their original ordering in the corresponding \apr{} tools. Then, \tech{} executes the patch against the original test suite, and stores the patch execution results into the patch validation matrix \mathmv{} (Line 5). If \mathp{} is a plausible patch, it will be stored in the resulting set \plau{} for manual inspection (Lines 6-7). To help with on-the-fly patch prioritization, \tech{} computes the patch quality information for the current patch following Section~\ref{section:basic} (Line 8). Next, \tech{} goes through all the remaining patches to update their similarity tuples following Section~\ref{section:patch_similarity} (Lines 9-15). Note that all remaining patches will be compared with the newly executed patch to incrementally update their corresponding similarity tuples. Lastly, the priority scores for all remaining patches will be updated based on the updated similarity tuples following Section~\ref{section:patch_pri} (Line 16). In this way, the algorithm will proceed until all patches have been validated or the developers find a high-quality patch.

Note that the time complexity of the \tech{} algorithm is $O(n^2)$ at first glance ($n$ denotes the number of patches), since all the remaining patches need to be updated after each patch execution. Meanwhile, during our implementation, we realize that the similarity scores do not need to be updated for each remaining patch; instead, we can cluster all remaining patches based on the set of program elements they modify, since all patches with the same set of modified elements will have the same priority. In this way, the time complexity can be reduced to $O(nm)$, where $m$ denotes the number of patch clusters with the same modified element sets. Given $m<<n$ in practice, our actual \tech{} implementation incurs negligible overhead.

\input{paper-sections/definitions/algorithm}

\subsubsection{Example}
\label{section:example}

Let us now use the partial patch validation matrix\footnote{Note that we use partial since most \apr{} tools will collect partial matrices, but our idea generalizes to full matrices (as studied in Section~\ref{section:rq4}).} shown in Table~\ref{figure:quality-matrix} and its corresponding patch modification matrix (shown in Table~\ref{figure:similarity-elements}) as the example to illustrate our \tech technique. For this example, if we follow the original patch execution ordering (top-down), we need to execute four patches before finding the final plausible patch. Now we discuss how our \tech{} can help speed up this process.


Shown in Tables~\ref{figure:example-i1} and \ref{figure:example-i2}, Column ``\textit{Quality}'' describes the patches' actual quality (available after the corresponding patch validation);
Column ``\textit{Match}'' describes the number of matching elements against the last executed patch for each patch;
Column ``\textit{Differ}'' describes the number of differing elements against the last executed patch for each patch;
Columns ``\textit{\tp{}}'', ``\textit{\fp{}}'', ``\textit{\tn{}}'', and ``\textit{\fn{}}'' represent the \textit{accumulated} \tftuple{} per patch;\footnote{Note, each tuple is initialized with \tp{}=1, \fp{}=1, \tn{}=1, and \fn{}=1 for numerical stability.} lastly, Column ``\textit{Score}'' represents the Ochiai calculated priority score as defined in Equation (\ref{formula:Ochiai}).

In the first iteration (shown in Table~\ref{figure:example-i1}), \tech{} will compute the quality of the executed patch, \patcha{} (marked with gray). We can immediately determine that the patch is low quality simply because it cannot make any originally failing tests pass. Note that we also show the quality for all other unexecuted patches to illustrate the quality computation. Then, given \patcha{} has been executed, we can update the similarity tuple for each remaining patch. For example, for \patchb{}, the set intersection and symmetrical set difference with \patcha{} is 
\{\elea{}, \eleb{}, \elec{}\} $\cap$ \{\elea{}, \eleb{}, \elec{}, \eled{}\} = \{\elea{}, \eleb{}, \elec{}\} 
and 
\{\elea{}, \eleb{}, \elec{}\} \setdiff{} \{\elea{}, \eleb{}, \elec{}, \eled{}\} = \{\eled{}\}, respectively. 
Therefore, since \patcha{} is a low-quality patch, \tn{} increments by 3 and \fn{} increments by 1, resulting in the tuple (\tp{}=1+0, \fp{}=1+0, \tn{}=1+3, \fn{}=1+1). Similarly, we can compute the similarity tuples for all the other remaining patches as shown in Table~\ref{figure:example-i1}. Then, via applying the default Ochiai formula on the computed tuples, we can compute the priority scores for all the three remaining patches as shown in Column ``Score'' in Table~\ref{figure:example-i1}. In this way, the patch with the highest priority, \patchd{}, is selected for the next patch execution.

Then, in the second iteration, \patchd{} gets executed (marked in gray) as shown in Table~\ref{figure:example-i2}. Note that \patchd{} is a plausible patch that can pass all system tests. Therefore, the developers can immediately start the patch inspection to check if \patchd{} is the correct patch. Of course, the patch execution can still continue if \patchd{} is not the correct patch. Continuing the algorithm, the remaining patches will be further compared with the newly executed \patchd{} to update their similarity tuples. For example, for \patchb{}, the set intersection and symmetrical set difference with \patchd{} is 
\{\elea{}, \eleb{}, \elec{}, \eled{}\} $\cap$ \{\elea{}\} = \{\elea{}\} 
and 
\{\elea{}, \eleb{}, \elec{}, \eled{}\} \setdiff{} \{\elea{}\} = \{\eleb{}, \elec{}, \eled{}\}, respectively. 
Since \patchd{} is a high-quality patch, \patchb{}'s tuple is updated by incrementing \tp{} by 1 and \fp{} by 3, resulting in the tuple (\tp{}=1+1, \fp{}=1+3, \tn{}=4+0, \fn{}=3+0). In this way, we can compute the scores for the remaining patches to further continue with patch validation.

\input{tables/sec03/primary-example}


For this example, we observe that the original patch execution ordering requires 4 patch executions to find the first plausible patch, while our \tech{} reduces the number of required patch executions to only 2, i.e., $\frac{4-2}{4}=50\%$ \dr{}. In this way, the developers can start manual patch inspection as soon as after 2 patch validations, substantially speeding up the repair process.

\subsection{\tech{} Variants}
\label{section:variants}

Potential \tech{} configurations are described in Table~\ref{figure:configurations} (default configuration settings \textbf{bolded}):

\subsubsection{Patch-Prioritization Formulae.} Besides Ochiai, other SBFL formulae can also be applied here. In particular we study all popular SBFL formulae from prior work~\cite{zhang2017boosting} in Section \ref{section:rq2}.
\subsubsection{Validation-Matrix Types.} Besides the partial patch-validation matrices widely used in practice, we also consider the impact of full validation matrices on our approach, discussed in Section \ref{section:rq4}. 
\subsubsection{Modification-Matrix Granularities.} Patch modification matrices are applicable to various granularities with varying patch similarity results. We examine the impact of such granularities on \tech{} in Section \ref{section:rq3}.

\subsubsection{Additional Patch Similarity Information}
By default, \tech{} only uses the set of modified program elements to calculate patch similarity for prioritizing patches on-the-fly. Another \tech{} extension is to compute the similarity score with additional information in addition to just the modified elements. Therefore, we further study another \tech{} variant, \techPat{}, which additionally considers that patches sharing the same fixing patterns may also share similar program behaviors. In this way, we can promote patches applying the same fixing patterns with known high-quality patches to further boost \tech.

\input{paper-sections/definitions/def-modification-matrix-++}

More specifically, \techPat only slightly differs from the default \tech when computing patch similarity, e.g., \techPat considers both (1) the set of modified elements and (2) the applied repair patterns. Based on the above patch-pattern matrix definition, we can recompute the similarity tuples for further improving \tech, e.g., \tp{}[\mathp{}] in Equation~(\ref{equation:tp}) becomes:

\begin{equation}
\label{equation:tp++}
\begin{split}
\text{\tp{}[\mathp{}]} = & \text{$\sum{}_{\mathbf{p}'}${ | \{ \mathe{} | \mathe{} $\in$ \mathRow{\mathmm{}}{\mathp{}} $\cap$ \mathRow{\mathmm{}}{\mathp{}'} $\wedge$ \mathp{}' $\in$ \mathPh{} \}|}~+}
\\ 
& \text{$\sum{}_{\mathbf{p}'}${ | \{ \mathe{} | \mathe{} $\in$ \mathRow{\mathmp{}}{\mathp{}} $\cap$ \mathRow{\mathmp{}}{\mathp{}'} $\wedge$ \mathp{}' $\in$ \mathPh{} \}|}}
\end{split}
\end{equation}

\input{tables/sec03/configurations}

\subsection{Further Extension Considering \apr Results from Other Tools}
\label{section:extensions}

In practice one repair tool is often insufficient to successfully find a correct patch. Thus developers often run multiple repair tools, ideally using the results of previously executed tools to optimize future executed tools. Similarly, the execution results of \textbf{other tools on the same subject} can be used to more guide \tech{} towards high-quality patches sooner. In particular, we use the repair information from \textit{all but one} repair tool to initialize the priority score of all patches. For example, when applying \tech{} to \tbar{} + Chart-1, all other tools with Chart-1 execution results will follow the same process outlined in Section \ref{section:alg} to initialize the priority scores of all \tbar{} + Chart-1 patches. With the priority scores initialized, \tech{} starts with the most prioritized patch and follows the algorithm outlined in Algorithm \ref{alg:seapr}, updating the already initialized priority scores of each patch. 

%% file: paper-sections/definitions/def-validation-matrix.tex
\begin{theorem}
    \textbf{Patch validation matrix}: Matrix \mathmv{} defines the validation results of all tests against all patch candidates. 
    In the matrix, each cell describes the validation result of test \matht{} $\in$ \mathT{} against patch \mathp{} $\in$ \mathP{}.
    Possible values for each cell are as follows: (1) \skipped{} if \matht{} remains unvalidated, (2) \failed{} if \matht{} fails on \mathp{} and (3) \passed{} if \matht{} passes on \mathp{}.
    \label{definition-validation}
\end{theorem}

\input{tables/sec03/example-quality}

%% file: tables/sec03/example-quality.tex
\begin{table}[htb]
    \scalebox{0.9}{
        \centering
        \begin{tabular}{|c|c|c|c|}
            \hline
            \rowcolor{tablegray} Patch ID & t$_{1}$ & t$_{2}$ & t$_{3}$ \\
            \hline
            \hline \rowcolor{tablegray}
                \textbf{\mathpb{} (buggy ver.)} & \textbf{\failed{}} & \textbf{\passed{}} & \textbf{\failed{}} \\
                \hline
                \patcha{} & \failed{} & \failed{} & \failed{} \\
                \patchb{} & \failed{} & \failed{} & \failed{} \\
                \patchc{} & \passed{} & \failed{} & \passed{} \\
                \patchd{} & \passed{} & \passed{} & \passed{} \\
            \hline
        \end{tabular}
    }
    \hfill{} 
    \scalebox{0.9}{
        \centering
        \begin{tabular}{|c|c|c|c|}
            \hline
            \rowcolor{tablegray} Patch ID & t$_{1}$ & t$_{2}$ & t$_{3}$ \\
            \hline
            \hline \rowcolor{tablegray}
                \textbf{\mathpb{} (buggy ver.)} & \textbf{\failed{}} & \textbf{\passed{}} & \textbf{\failed{}} \\
                \hline
                \patcha{} & \failed{} & \skipped{} & \skipped{} \\
                \patchb{} & \failed{} & \skipped{} & \skipped{} \\
                \patchc{} & \passed{} & \failed{} & \skipped{} \\
                \patchd{} & \passed{} & \passed{} & \passed{} \\
            \hline
        \end{tabular}
    }
    \caption{Example of full and partial patch-validation matrix}
    \label{figure:quality-matrix}
\end{table}

%% file: paper-sections/definitions/def-modification-matrix.tex
\begin{theorem}
    \textbf{Patch modification matrix}: Matrix \mathmm{} presents all program elements modified within each patch.
    Each cell describes if patch \mathp{} $\in$ \mathP{} modifies element \mathe{} $\in$ \mathE{} (i.e., all possible program elements).
    Acceptable values for each cell are as follows: (1) \passed{} if \mathp{} modifies element \mathe{} and (2) \skipped{} if \mathp{} does not modify element \mathe{}.
    \label{definition-modification}
\end{theorem}

\input{tables/sec03/similarity-elements}

%% file: tables/sec03/similarity-elements.tex
\begin{table}[htb]
    \scalebox{0.9}{
    \centering
    \begin{tabular}{|c||c|c|c|c||c|}
        \hline
    
        
        \Comment{\hhline{
            |
            >{\arrayrulecolor{tablegray}}-
            >{\arrayrulecolor{black}}||
            *{4}{>{\arrayrulecolor{tablegray}}->{\arrayrulecolor{black}}|}
            >{\arrayrulecolor{black}}|-|
        }}
        
        \rowcolor{tablegray} \multirow{-1}{*}{\textbf{Patch ID}} & \multirow{-1}{*}{\textbf{\elea{}}} & \multirow{-1}{*}{\textbf{\eleb{}}} & \multirow{-1}{*}{\textbf{\elec{}}} & \multirow{-1}{*}{\textbf{\eled{}}} & \textbf{Modified Element(s)} \\
    
        \hline
        \hline
        
        \patcha{} & \checkmark{} & \checkmark{} & \checkmark{} & \skipped{} & \{\elea{}, \eleb{}, \elec{}\} \\
        \patchb{} & \checkmark{} & \checkmark{} & \checkmark{} & \checkmark{} & \{\elea{}, \eleb{}, \elec{}, \eled{}\} \\
        \patchc{} & \skipped{} & \checkmark{} & \checkmark{} & \skipped{} & \{\eleb{}, \elec{}\} \\
        \patchd{} & \checkmark{} & \skipped{} & \skipped{} & \skipped{} & \{\elea{}\} \\
            
        \hline    
    \end{tabular}
    }
    \caption{Example of patch modification matrix}
    \label{figure:similarity-elements}
\end{table}

%% file: paper-sections/definitions/algorithm.tex

\begin{algorithm}[t!]
\small
    \SetKwProg{BeginFunc}{function}{:}{end}
    \SetKwProg{try}{try}{:}{}
    \SetKwProg{catch}{catch}{:}{end}
    \caption{\label{alg:seapr} \tech Algorithm}
    \KwIn{The original buggy program \mathpb, test suite \mathT, the list of candidate patches \mathP, the similarity tuples \tp, \tn, \fp, \fn}
    \KwOut{Plausible patches: \plau}
    \Begin{
        Initialize \tp, \tn, \fp, \fn \\
        \While{\mathP{} is not empty}{
            \mathp{} $\leftarrow$ $\mathtt{pop}$(\mathP) \tcp*[l]{pop the remaining patch with the highest priority} 
            \mathmv{} $\leftarrow$ $\mathtt{execute}$(\mathp, \mathT)\tcp*[l]{ get the execution results}
            \If{\mathp{} \text{is} $\mathtt{PLAUSIBLE}$}{
                    \plau $\leftarrow$ \mathp \tcp*[l]{put \mathp{} into the resulting set for manual inspection if it is plausible} 
            }
            
            \res{} $\leftarrow$ $\mathtt{computePatchQuality}$(\mathp, \mathpb, \mathmv)\tcp*[l]{compute the patch quality} 
            
            \tcp{Incrementally update the \tftuple{} for the remaining patches based the newly executed patch}
            
            \For{\mathp' $\in$ \mathP}{
                \If{\res{} = $\mathtt{HIGH-QUALITY}$}{
                    \tp[\mathp'] += |\mathRow{\mathmm{}}{\mathp{}} $\cap$ \mathRow{\mathmm{}}{\mathp{}'}|\\
                    \fp[\mathp'] += |\mathRow{\mathmm{}}{\mathp{}} \setdiff{} \mathRow{\mathmm{}}{\mathp{}'}|
                }
                \If{\res{} = $\mathtt{LOW-QUALITY}$}{
                    \tn[\mathp'] += |\mathRow{\mathmm{}}{\mathp{}} $\cap$ \mathRow{\mathmm{}}{\mathp{}'}|\\
                    \fn[\mathp'] += |\mathRow{\mathmm{}}{\mathp{}} \setdiff{} \mathRow{\mathmm{}}{\mathp{}'}|
                }
            }
            \mathP $\leftarrow$ $\mathtt{computePriorityScore}$(\mathP, \tp, \fp, \tn, \fn)\tcp*[l]{update priority scores for remaining patches} 
        }
    }
\end{algorithm}

%% file: tables/sec03/primary-example.tex

\begin{table}[tbh]
    \scalebox{0.78}{
    \centering
        \begin{tabular}{|c|c||c|c||c|c|c|c||c|}
            \hline
            \rowcolor{tablegray} \textbf{ID} & Quality & Match & Differ & \tp{} & \fp{} & \tn{} & \fn{} & \textbf{Score} \\
            \hline
            \hline
            
            \rowcolor{tablegray} \textbf{\patcha{}} & \textbf{Low} & \textbf{-} & \textbf{-} & \textbf{-} & \textbf{-} & \textbf{-} & \textbf{-} & \textbf{-} \\
            \hline
            \patchb{} & Low & \{\elea{}, \eleb{}, \elec{}\} & \{\eled{}\} & 1+0 & 1+0 & 1+3 & 1+1 & 0.32 \\
            \patchc{} & High & \{\eleb{}, \elec{}\} & \{\elea{}\} & 1+0 & 1+0 & 1+2 & 1+1 & 0.35 \\
            \patchd{} & Plausible & \{\elea{}\} & \{\eleb{}, \elec{}\} & 1+0 & 1+0 & 1+1 & 1+2 & 0.50 \\
            \hline    
        \end{tabular}
    }
    
    \caption{Detailed \tech{} step-by-step when processing \patcha{}}
    \label{figure:example-i1}
\end{table}


\begin{table}[tbh]
    \scalebox{0.78}{
    \centering 
        \begin{tabular}{|c|c||c|c||c|c|c|c||c|}
            \hline
            \rowcolor{tablegray} \textbf{ID} & Quality & Match & Differ & \tp{} & \fp{} & \tn{} & \fn{} & \textbf{Score} \\
            \hline
            \hline
            
            \rowcolor{tablegray} \patcha{} & Low & - & - & - & - & - & - & - \\
            \rowcolor{tablegray} \textbf{\patchd{}} & \textbf{Plausible} & \textbf{-} & \textbf{-} & \textbf{-} & \textbf{-} & \textbf{-} & \textbf{-} & \textbf{-} \\
            \hline
            \patchb{} & Low & \{\elea{}\} & \{\eleb{}, \elec{}, \eled{}\} & 1+1 & 1+3 & 4+0 & 2+0 & 0.33 \\
            \patchc{} & High & \{\eleb{}, \elec{}\} & \{\elea{}, \eled{}\} & 1+2 & 1+2 & 3+0 & 2+0 & 0.50 \\
            \hline    
        \end{tabular}
    }
    
    \caption{Detailed \tech{} step-by-step when processing \patchd{}}
    \label{figure:example-i2}
\end{table}

%% file: paper-sections/definitions/def-modification-matrix-++.tex
\begin{theorem}
    \textbf{Patch repair pattern matrix}: Matrix \mathmp{} presents the applied repair patterns applied to each patch.
    Each cell describes if patch \mathp{} $\in$ \mathP{} applies repair pattern \mathr{} $\in$ \mathR{} (i.e., all predefined repair patterns).
    Acceptable values for each cell are as follows: (1) \checkmark{} if \mathp{} applies pattern \mathr{} and (2) \skipped{} if \mathp{} does not apply pattern \mathr{}.
    \label{definition-pattern}
\end{theorem}

\input{tables/sec03/similarity-patterns}

%% file: tables/sec03/similarity-patterns.tex
\begin{table}[htb]
    \scalebox{0.9}{
    \centering
    \begin{tabular}{|c||c|c|c||c|c|}
        \hline
    
        \rowcolor{tablegray} & & & & \multicolumn{2}{c|}{\textbf{\tech{} Features}} \\ 
            \hhline{
            |
            >{\arrayrulecolor{tablegray}}-
            >{\arrayrulecolor{black}}||
            *{3}{>{\arrayrulecolor{tablegray}}->{\arrayrulecolor{black}}|}
            >{\arrayrulecolor{black}}|--|
            }
        \rowcolor{tablegray} \multirow{-2}{*}{\textbf{Patch ID}} & \multirow{-2}{*}{\textbf{\pata{}}} & \multirow{-2}{*}{\textbf{\patb{}}}  & \multirow{-2}{*}{\textbf{\patc{}}}  & \textbf{Modified Element(s)} & \textbf{Pattern(s)} \\
    
        \hline
        \hline
        
        \patcha{} & \checkmark{} & \skipped{} & \skipped{} & \{\elea{}, \eleb{}, \elec{}\} & \{\pata{}\} \\
        \patchb{} & \skipped{} & \checkmark{} & \skipped{} & \{\elea{}, \eleb{}, \elec{}, \eled{}\} & \{\patb{}\} \\
        \patchc{} & \skipped{} & \skipped{} & \checkmark{} & \{\eleb{}, \elec{}\} & \{\patc{}\}  \\
        \patchd{} & \skipped{} & \checkmark{} & \skipped{} & \{\elea{}\} & \{\patb{}\} \\
            
        \hline    
    \end{tabular}
    }
    \caption{Example of patch repair pattern matrix}
    \label{figure:similarity-patterns}
\end{table}

%% file: tables/sec03/configurations.tex
\begin{table}[htb]
    \scalebox{1.0}{
        \centering
       \begin{tabular}{|c|p{13em}|}
            \hline
            \rowcolor{tablegray} Parameter & Variant Settings \\
            \hline
            \hline
            
            Prioritization Formulae & Tarantula, \textbf{Ochiai}, Ochiai2, Op2, SBI, Jaccard, Kulczynski, Dstar2\\
            \hline
           \rowcolor{tablegray} Validation Matrices & Full Matrix, \textbf{Partial Matrix} \\
             \hline
             Modification Granularities & Package, Class, \textbf{Method}, Stmt \\
             \hline
            \rowcolor{tablegray} Similarity Computation & \textbf{Modified Elements}, Modified Elements+Patch Patterns \\
            \hline
        \end{tabular}
    }
    \caption{Possible configurations for \tech{}}
    \label{figure:configurations}
\end{table}

%% file: paper-sections/04_study-design.tex
\section{Experimental Design}
\label{section:design}
\subsection{Research Questions}

In this paper, we aim to study the following research questions to thoroughly evaluate the performance of our \tech{}:
\begin{itemize}
\item \textbf{RQ1:} How does the default \tech{} perform on state-of-the-art \apr{} systems?
\item \textbf{RQ2:} How do different formulae affect the effectiveness of \tech{}?
\item \textbf{RQ3:} How do different patch-modification-matrix granularities affect the effectiveness of \tech{}?
\item \textbf{RQ4:} How does patch validation matrices impact \tech{}?
\item \textbf{RQ5:} How does \tech{} perform when leveraging additional patch similarity information?
\item \textbf{RQ6:} How does \tech{} perform when further leveraging historical repair information from other \apr{} systems?
\end{itemize}

Note that we first study our default \tech{} configuration with the Ochiai formula, method granularity, partial patch validation matrices, modified elements similarity computation in RQ1. Then, in RQs 2-5, we investigate the impact of different formulae, granularities, types of patch validation matrices, as well as the usage of additional patch execution information. Lastly, in RQ6, we investigate the effectiveness of \tech{} when further leveraging patch validation information from other \apr{} tools.

\subsection{Evaluation Dataset}
\label{section:evaluation-dataset}

We choose to evaluate the proposed technique against the \dfts{} dataset (V1.2.0), since \dfts{} (V1.2.0) is the most widely used \apr{} dataset to date, and will allow our technique to be easily compared to and replicated in the future. The details for our adopted dataset are shown in Table \ref{table:defects-composition}. In the table, Column ``\textit{\# Bugs}'' presents the number of buggy versions studied for each subject. Column ``\textit{\# Tests}'' and Column ``\textit{LOC}'' presents the number of JUnit tests and lines of code available within the head (i.e., most recent) version of each subject, respectively. 

\input{tables/sec04/defect4j-composition}

\subsection{Tools}
\label{section:tools}
Following prior work~\cite{Liu2020OnPrograms\Comment{, Benton2020}}, we consider all \toolsConsidered{} program repair tools publicly available and applicable to \dfts{} in this study, show in Table \ref{table:tools}. Of these tool candidates, we exclude \acs{}, \dynamoth{}, and \nopol{} due to low patch generation (i.e., $<$500 total patches across all studied \dfts{} projects in total\Comment{$<$ 1 patch per subject density}). Evaluating our technique on such tools with small number of patches will induce noise into our results; in addition, such \apr{} tools do not need patch prioritization anyhow\checkLater{ \sam{(1) statement seems to conflict with our motivation that apr tools are slow unless we include some time information on these tools}}. We further exclude \simfix{} since the tool stops execution after finding the first plausible patch. The validation results of such tools are undegradable, since the last patch is always the desirable plausible patch. Results from such tools can only improve, biasing our findings. With the exclusion of these tools, this leaves us with \toolsActual{} repair tools used in this paper.

\input{tables/sec04/tools}

\subsection{Evaluation Metrics}

We have adopted the following two evaluation metrics: the reduction on the number of patch executions before finding the first (1) \textbf{plausible} patch and (2) \textbf{correct} patch. We study the number of patches to the first plausible patch since in practice developers will begin patch inspection after the first plausible patch is found; in this way, faster plausible patch detection can enable developers to start patch inspection earlier (and potentially speed up the \apr{} process). Similarly, we study the number of patches to the first correct patch since developers will stop the patch validation process once the correct patch is found; in this way, faster correct patch detection can save overall \apr{} time.
Also note that we leverage the reduction on the number of patch executions as recommended by prior work~\cite{Liu2020OnPrograms} since time cost information tends to introduce bias: it is dependent on many factors (e.g., specific implementations and test execution engines) unrelated to APR approaches and is often unstable. 

To this end, Formula \ref{formula:displacement-ratio} represents our primary evaluation metric. 
\textbf{\baselineOld{}} represents the position of the first desirable (i.e., plausible or correct) patch, pre-priortization.
\textbf{\baselineNew{}} represents position of the first desirable patch, post-prioritization.
Note that when multiple desirable patches are produced, the initial desireable patch and the final desireable patch are not necessarily the same patch.

\begin{equation}
  Reduction = \dfrac{P_{baseline} - P_{new}}{P_{baseline}}
  \label{formula:displacement-ratio}
\end{equation}

\subsection{Experimental Procedure}
For each studied \apr{} tool, we evaluate the effectiveness of \tech{} on all patches that can be generated and validated by the tool within its original time limit. 
We first analyzed the original execution of each tool on a subject-by-subject basis to obtain (1) the original patch execution ordering per repair tool and (2) the position of the earliest plausible/correct patch. 

After information collection, we then repeat the patch validation process for each tool again with our \tech{} on-the-fly patch prioritization. Note that due to the large effort required to modify all used tools, we choose to simulate the validation process of each tool instead of modifying each tool directly.
For each given subject, \tech{} initially executes the \textbf{first} patch produced by the tool. After the first patch execution, \tech{} iterates through all patches not yet validated, following Algorithm~\ref{alg:seapr}, to record the new position for the first plausible/correct patch.

All our experiments were conducted within the following environment: 36 3.0GHz Intel Xeon Platinum Processors, 60GBs of
memory, and Ubuntu 18.04.4 LTS operating system.


%% file: tables/sec04/defect4j-composition.tex
\begin{table}[tbh]
    \scalebox{0.95}{
        \centering
        \begin{tabular}{|r|r|c|c|c|} 
            \hline
            \rowcolor{tablegray} \textbf{Subject} & \textbf{Name} & \textbf{\# Bugs} & \textbf{\# Tests} & \textbf{LOC} \\
            \hline
            \hline
            Chart & JFreeChart & 26 & 2,205 & 96K \\
            \rowcolor{tablegray}Lang & Apache Lang & 65 & 2,245 & 22K \\
            Math & Apache Math & 106 & 3,602 & 85K \\
            \rowcolor{tablegray}Time & Joda-Time & 27 & 4,130 & 28K \\
            Mockito & Mockito framework & 38 & 1,366 & 23K \\
            \rowcolor{tablegray}Closure & Google Closure compiler & 133 & 7,927 & 90K \\
            \hline
            \hline
            \multicolumn{2}{|r|}{\textbf{Total}} & \textbf{395} & \textbf{21,475} & \textbf{344K} \\
            \hline
        \end{tabular}
    }
    \caption{Studied bugs from \dfts{} v1.2.0}
    \label{table:defects-composition}
\end{table}

%% file: tables/sec04/tools.tex
\begin{table}[tbh]
\scalebox{1.0}{
    \centering
    \begin{tabular}{|r|p{17em}|}
    
        \hline
        
        \rowcolor{tablegray}
         
        \textbf{Tool Category} & \textbf{Tool(s)} \\
        \hline
        \hline
        
        Constraint-based & \sout{\acs{}}~\cite{Xiong2017PreciseRepair}, \cardumen{}~\cite{toolCardumen}, \sout{\dynamoth{}}~\cite{toolDynamoth}, \sout{\nopol{}}~\cite{toolNopol} \\ \rowcolor{tablegray}
        
        Heuristic-based & \arja{}~\cite{toolArja}, \genprog{}~\cite{toolArja}, \jgenprog{}~\cite{toolAstor}, \jkali{}~\cite{toolAstor}, \jmutrepair{}~\cite{toolAstor}, \kali{}~\cite{toolArja}, \rsrepair{}~\cite{toolArja}, \sout{\simfix{}}~\cite{Jiang2018ShapingCode} \\
        
        Template-based & \avatar{}~\cite{toolAvatar}, \fixminer{}~\cite{toolFixminer}, \kpar{}~\cite{Liu2019YouSystems}, \tbar{}~\cite{toolTbar} \\
         
         \hline
    \end{tabular}}
    \caption{Repair tools under consideration}
    \label{table:tools}
\end{table}


%% file: paper-sections/05_analysis.tex
\section{Result Analysis}
\label{section:analysis}

\input{paper-sections/research-questions/rq1/content}
\input{paper-sections/research-questions/rq2/content}
\input{paper-sections/research-questions/rq3/content}
\input{paper-sections/research-questions/rq4/content}
\input{paper-sections/research-questions/rq6/content}
\input{paper-sections/research-questions/rq5/content}

\checkLater{
\subsection{Threats to Validity}
\subsubsection{Internal Validity}

All of our results are dependent on the correctness of our implementation of all the studied techniques. We mitigate this threat, by using existing data publicly distributed by \profl{} and \unidebug{} authors \cite{unidebugWeb}. We further perform a peer review of our metric collection and evaluation scripts to ensure accurate and correct results.

\subsubsection{External Validity}

While our approach is generalizable to any type of patch-and-validate system, we only evaluate Java-based \apr{} tools which may skew results. To mitigate this threat we 1) used a wide variety of \apr{} tools (e.g. Constraint-based, Heuristic-based, and Template-based) and 2) consider tools actively used in recent and related works. We also actively evaluate our technique on a recent version of \dfts{}, containing a wide variety of real-world projects.

\subsubsection{Construct Validity}

A major threat to construct validity lies in our choice of evaluation metrics. To mitigate this threat, (1) we thoroughly describe our primary evaluation metrics.

}

%% file: paper-sections/research-questions/rq1/content.tex
\subsection{RQ1: Overall \tech{} Effectiveness}
\label{section:rq1}

\noindent \textbf{Quantitative Analysis.}
In this section, we first investigate the overall effectiveness of our default \tech{} on all \toolsActual{} studied repair tools against all the six studied subjects from the \dfts{} benchmark. Table~\ref{table:rq1} shows the patch reduction in terms of the first \textbf{plausible patches}\Comment{\lingming{let's just call it the patch reduction in terms of the first plausible/correct patch; change globally}\Xia{DONE}} with the default configuration settings.
In this table, each row and column represent different \dfts{} projects and repair tools, respectively.
Note that each cell presents the overall \dr{} for each tool on the corresponding project 
with symbol ``---'' indicating that the repair tool cannot generate any plausible patches on the corresponding project. 
We also present the overall \dr{}s on all projects, shown in the last row. 
From this table, we have the following observations. First, \tech{} can improve the overall effectiveness of patch validations for almost all repair tools. 
For example, \tech{} can reduce patch validation by 78.81\% on \genprog{} and 54.80\% on \avatar{}. 
Meanwhile, \tech degrades the performance of \cardumen{}\Comment{ and \jgenprog{}} by 7.32\%\Comment{\lingming{it should be \tech reduces the performance of \cardumen{} and \jgenprog{} by...}}. 
The possible reason is that \cardumen{} only generate plausible patches for very few buggy versions\checkLater{\lingming{why the cardumen paper claims it can produce 8935 plausible patches for 77 buggy versions?} \sam{they claim they can generate \textbf{8.9k patches after running the tool 10x on each subject}}\lingming{does cardumen collect full or partial matrix bug default?}}, and thus the overall results can have large variance. 
Second, for most repair tools, \tech{} can also improve patch validation results for the majority of \dfts{} projects. For example, \tech can achieve positive reductions for \avatar{}, \fixminer{}, \genprog{}, \jkali{}, \jmutrepair{} and \kali{} on all the projects with plausible patches. For other repair tools, \tech{} degrades the performance of them on at most one or two projects. This finding shows the applicability of \tech{} on different repair tools and different projects.

Similarly, Table~\ref{table:rq1Correct} further shows the patch reduction ratios in terms of the first \textbf{correct patch}. 
Note that there exists no information on the number of correct patches reported in the original or subsequent papers for \genprog{}, \rsrepair{}, and \kali{}~\cite{toolArja}. 
For \jkali{} and \jgenprog{}, we cannot generate any correct patches as reported by the tool's original paper~\cite{toolAstor}, likely due to the tool's nondeterminism and different execution environments. Therefore, we exclude these five repairs tools from Table \ref{table:rq1Correct}. 
From the table, we can find that \tech{} can also substantially reduce patch validation (until the first correct patch) for most repair tools. For example, \tech{} can improve the overall performance of \arja{}, \avatar{}, \jmutrepair{}, \kpar{} and \tbar{} by 72.96\%, 80.15\%, 53.57\%, 64.06\% and 48.80\% respectively. Another interesting finding is that \tech{} does not degrade the performance of any of the studied tools, although it cannot reduce patch executions for some repair tools (i.e., 0\% reduction). The possible reason may be that the original ranking of correct patches by these tools is good enough, leaving limited room of improvement for our \tech{}.\checkLater{\lingming{not convincing, if it was good enough why \tech{} is not worse? also, put the ranking of correct patches here may help}}

Note that since patch reductions in terms of correct patches do not apply to all studied subjects and are mainly consistent with that in terms of plausible patches, we use patch reductions in terms of plausible patches for our following RQs. 

\input{paper-sections/research-questions/rq1/table}

\input{paper-sections/research-questions/rq1/table-correct}

\noindent \textbf{Qualitative Analysis.}
Next, we perform detailed examples to investigate the performance of \tech{}.

\input{paper-sections/research-questions/rq1/example-single-edit-demotion}
\input{paper-sections/research-questions/rq1/example-single-edit-promotion}
\input{paper-sections/research-questions/rq1/example-multi-edit-promotion}

\begin{finding}{}
    \tech{} can substantially reduce patch executions before finding the first plausible/correct patches for almost all studied repair tools, with a maximum improvement of 78.81\%/72.96\%.
\end{finding}

%% file: paper-sections/research-questions/rq1/table.tex
\begin{table*}[tbh]
    \scalebox{0.85}{
        \centering
        \begin{tabular}{|c|c|c|c|c|c|c|c|c|c|c|c|c|}
            \hline
            \rowcolor{tablegray} & \arja{} & \avatar{} & \cardumen{} & \fixminer{} & \genprog{} & \jgenprog{} & \jkali{} & \jmutrepair{} & \kali{} & \kpar{} & \rsrepair{} & \tbar{} \\
            \hline
            \hline
            
            Chart & 80.10\% & 79.36\% & -15.15\% & 48.49\% & 44.75\% & 3.33\% & 0.00\% & 11.36\% & 14.29\% & 21.27\% & 32.04\% & 48.61\% \\
            \rowcolor{tablegray} Closure & --- & 60.62\% & --- & --- & --- & --- & --- & --- & --- & 46.89\% & --- & 9.26\%  \\
            Lang & -836.92\% & 48.29\% & 32.00\% & 53.80\% & 57.45\% & -75.00\% & 48.15\% & 24.10\% & --- & -3.47\% & -13.00\% & -61.49\% \\
            \rowcolor{tablegray} Math & 19.62\% & --- & -37.50\% & --- & 81.25\% & 23.33\% & 38.10\% & 17.71\% & 3.99\% & 14.69\% & 28.82\% & 45.51\% \\
            Mockito & --- & 25.00\% & --- & --- & --- & --- & --- & --- & --- & 37.91\% & --- & 5.13\% \\
            \rowcolor{tablegray} Time & --- & 40.67\% & --- & 14.54\% & --- & --- & --- & --- & --- & 17.26\% & --- & -174.47\% \\

            \hline
            \hline
 \textbf{Overall} & \textbf{40.05\%} & \textbf{54.80\%} & \textbf{-7.32\%} & \textbf{43.55\%} & \textbf{78.81\%} & \textbf{7.81\%} & \textbf{37.18\%} & \textbf{20.26\%} & \textbf{6.36\%} & \textbf{17.56\%} & \textbf{21.84\%} & \textbf{10.93\%} \\
            \hline
            
        \end{tabular}
        }
        \caption{\tech{} default configuration results for plausible patches}
        \label{table:rq1}
\end{table*}

%% file: paper-sections/research-questions/rq1/table-correct.tex
 \begin{table}[tbh]
    \scalebox{0.73}{
        \centering
        \begin{tabular}{|c|c|c|c|c|c|c|c|c|c|c|c|c|}
        
            \hline
            \rowcolor{tablegray}  & \arja{} & \avatar{} & \cardumen{} & \fixminer{} & \jmutrepair{} & \kpar{} & \tbar{} \\
            \hline
            \hline
            
            Chart & --- & --- & 0.00\% & 0.00\% & --- & 84.97\% & 0.00\% \\
            \rowcolor{tablegray} Closure & --- & 0.00\% & --- & --- & --- & 0.00\% & 0.00\% \\
            Lang & --- & 93.86\% & --- & 0.00\% & --- & --- & 0.00\% \\
            \rowcolor{tablegray} Math & 72.96\% & --- & --- & --- & 53.57\% & 22.09\% & 65.16\% \\
            Mockito & --- & 0.00\% & --- & --- & --- & --- & 33.33\% \\
            \rowcolor{tablegray} Time & --- & --- & --- & --- & --- & --- & --- \\
            
            \hline
            \hline
            \textbf{Overall} & \textbf{72.96\%} & \textbf{80.15\%} & \textbf{0.00\%} & \textbf{0.00\%} & \textbf{53.57\%} & \textbf{64.06\%} & \textbf{48.80\%} \\
            \hline
        \end{tabular}
        }
        \caption{\tech{} default configuration results for correct patches\Comment{\lingming{add the detailed reason, and remove columns for those tools without correct patches to make the table to be single-column} \mengshi{Table corrected}}}
        \label{table:rq1Correct}
\end{table}

%% file: paper-sections/research-questions/rq1/example-single-edit-demotion.tex
\begin{figure}
\centering
    \begin{lstlisting}[backgroundcolor=\color{tablegray},language=JAVA,basicstyle=\ttfamily\tiny,frame=single,numbers=left]
// org.jfree.chart.plot.CategoryPlot.java
@@ -911,7 +911,6 @@
      * @param axis  the axis.
      */
     public void setRangeAxis(int index, ValueAxis axis) {
-        setRangeAxis(index, axis, true);
     }

     /**
    \end{lstlisting}
    \caption{\arja{} Chart-19 prioritizer patch}
    \label{code:arja-C19-prioritizer}
\end{figure}

\begin{figure}
\centering
    \begin{lstlisting}[backgroundcolor=\color{tablegray},language=JAVA,basicstyle=\ttfamily\tiny,frame=single,numbers=left]
// org.jfree.chart.uti.AbstractObjectList.java
@@ -161,7 +161,10 @@
                 return (index);
             }
         }
-        return -1;
+        if (object == null) {
+            throw new IllegalArgumentException("Null 'object' argument.");
+       }
+       return -1;
     }
     
     /**
    \end{lstlisting}
    \caption{\arja{} Chart-19 plausible patch}
    \label{code:arja-C19-plausible}
\end{figure}

\subsubsection{Example 1}
Figure \ref{code:arja-C19-prioritizer} shows one of the low-quality patches produced by \arja{} on Chart-19 while Figure \ref{code:arja-C19-plausible} shows one of few generated plausible patches. Note that these two patches modify different files. In this example, we observe how low-quality patches can help prioritizing plausible patches. According to our technique, other patches \textit{sharing similar modified elements} with these low-quality patches are \textbf{deprioritized}. Thus as any low-quality patch modifying one particular set of methods is validated, all other similar patches are deprioritized. This essentially results in (1) the clustering of patches based on the set of modified methods and (2) prioritizing / deprioritizing these clusters. Upon validation of consecutive low-quality patches, this phenomenon culminates to a breadth-first exploration of the patch clusters, mitigating the risk of some high-quality patches getting starved. This process repeats until \arja{} finds plausible patch from Figure \ref{code:arja-C19-plausible}. For \arja{} and Chart-19, we observe \arja{} validates the plausible patch 33rd instead of 626th, achieving a \dr{} of 94.7\%.

%% file: paper-sections/research-questions/rq1/example-single-edit-promotion.tex
\begin{figure}
\centering
    \begin{lstlisting}[backgroundcolor=\color{tablegray},language=JAVA,basicstyle=\ttfamily\tiny,frame=single,numbers=left]
@@ -1621,9 +1621,9 @@ public static double distance(double[] p1, double[] p2) {    
* @return the L<sub>2</sub> distance between the two points            
*/            
public static double distance(int[] p1, int[] p2) {            
int sum = 0;            
for (int i = 0; i < p1.length; i++) {            
    final int dp = p1[i] - p2[i];            
-   sum += dp * dp;            
}            
return Math.sqrt(sum);
    \end{lstlisting}
    \caption{\tbar{} Math-79 prioritizer patch}
    \label{code:tbar-M79-prioritizer}
\end{figure}

\begin{figure}
    \centering
    \begin{lstlisting}[backgroundcolor=\color{tablegray},language=JAVA,basicstyle=\ttfamily\tiny,frame=single,numbers=left]
@@ -1621,9 +1621,9 @@ public static double distance(double[] p1, double[] p2) {    
* @return the L<sub>2</sub> distance between the two points            
*/            
public static double distance(int[] p1, int[] p2) {            
- int sum = 0;            
+ double sum = 0;            
  for (int i = 0; i < p1.length; i++) {            
-     final int dp = p1[i] - p2[i];            
+     final double dp = p1[i] - p2[i];            
      sum += dp * dp;            
}            
return Math.sqrt(sum);
    \end{lstlisting}
\caption{\tbar{} Math-79 plausible patch}
\label{code:tbar-M79-plausible}
\end{figure}

\subsubsection{Example 2}
Figure \ref{code:tbar-M79-prioritizer} presents a \textit{non-correct high-quality patch} produced by \tbar{} which immediately helps prioritize a \textit{plausible patch}, shown in Figure \ref{code:tbar-M79-plausible}. From these figures, it is evident that both patches modify method \texttt{distance(int[], int[])}. Upon processing the non-correct high-quality patch, the plausible patch in Figure \ref{code:tbar-M79-plausible} is immediately prioritized earlier by \tech{} since they share similar modified program elements. 
In \tbar{}, the plausible patch shown in Figure \ref{code:tbar-M79-plausible} is originally validated at the 125th place without our \tech{} technique, and now validated at the 33rd position, achieving a reduction of 73.6\%.

%% file: paper-sections/research-questions/rq1/example-multi-edit-promotion.tex
\begin{figure}
\centering
    \begin{lstlisting}[backgroundcolor=\color{tablegray},language=JAVA,basicstyle=\ttfamily\tiny,frame=single,numbers=left]
// org.jfree.chart.JFreeChart.java
@@ -1484,7 +1484,6 @@
      */
     public void fireChartChanged() {
         ChartChangeEvent event = new ChartChangeEvent(this);
-        notifyListeners(event);
     }

     /**
    \end{lstlisting}
    \caption{\arja{} Chart-12 prioritizer patch}
    \label{code:arja-C12-prioritizer}
\end{figure}

\begin{figure}
    \begin{subfigure}[b]{0.45\textwidth}
    \begin{tabular}{ll}
    \begin{lstlisting}[backgroundcolor=\color{tablegray},language=JAVA,basicstyle=\ttfamily\tiny,frame=single,numbers=left]
// org.jfree.data.general.AbstractDataset.java
@@ -158,7 +158,7 @@
      */
     public boolean hasListener(EventListener listener) {
         List list = Arrays.asList(this.listenerList.getListenerList());
-        return list.contains(listener);
+        return true;
     }

     /**
    \end{lstlisting}
    \end{tabular}
    \end{subfigure}


    \begin{subfigure}[b]{0.35\textwidth}
    \begin{tabular}{ll}
    \begin{lstlisting}[backgroundcolor=\color{tablegray},language=JAVA,basicstyle=\ttfamily\tiny,frame=single,numbers=left]
// org.jfree.chart.title.LegendTitle.java
@@ -539,7 +539,9 @@

      */
     public boolean equals(Object obj) {
         if (obj == this) {
-            return true;   
+            if (!super.equals(obj)) {
+                               return false;
+           }   
         }
         if (!(obj instanceof LegendTitle)) {
             return false;
    \end{lstlisting}
    \end{tabular}
    \end{subfigure}


    \begin{subfigure}[b]{0.40\textwidth}
    \begin{tabular}{ll}
    \begin{lstlisting}[backgroundcolor=\color{tablegray},language=JAVA,basicstyle=\ttfamily\tiny,frame=single,numbers=left]
// org.jfree.chart.JFreeChart.java
@@ -641,7 +641,10 @@

         int seen = 0;
         Iterator iterator = this.subtitles.iterator();
         while (iterator.hasNext()) {
-            Title subtitle = (Title) iterator.next();
+            if (this.title != null) {
+                               this.title.addChangeListener(this);
+                       }
+                       Title subtitle = (Title) iterator.next();
             if (subtitle instanceof LegendTitle) {
                 if (seen == index) {
                     return (LegendTitle) subtitle;
    \end{lstlisting}
    \end{tabular}
    \end{subfigure}
    \caption{\arja{} Chart-12 plausible patch}
    \label{code:arja-C12-plausible}
\end{figure}

\subsubsection{Example 3}
We now show the effectiveness of \tech{} on multi-edit APR techniques (e.g. \arja{}). Figure \ref{code:arja-C12-prioritizer} presents one of the 60+ low-quality patches generated by \arja{} on subject Chart-12 with each patch potentially modifying one or more code elements. Figure~\ref{code:arja-C12-plausible} shows the first plausible patch. From the two figures, it is evident that the plausible patch and this particular low-quality patch do not modify the same code elements. In fact, despite modifications across multiple files, \tech{}'s performance does not change. 
That is, the validation of these numerous low-quality patches deprioritizes all other patches in the same cluster. As \tech{} iterates through these clusters, a plausible patch is quickly found shown in Figure \ref{code:arja-C12-plausible}. This phenomenon leads \tech{} to prioritize the first plausible patch of Chart-12 from 616th to 63rd, an improvement of 89.8\%.

%% file: paper-sections/research-questions/rq2/content.tex
\subsection{RQ2: Impact of Different Formulae}
\label{section:rq2}

\input{paper-sections/research-questions/rq2/table}


In the Section ~\ref{section:rq1}, we studied \tech{} with the default \ochiai{} formula. In fact, \tech{} is a general approach and can leverage any other existing SBFL formula. 
Therefore, in this section, we further investigate the impact of different SBFL formulae on the effectiveness of \tech{}. 
Table ~\ref{table:rq2} shows the experimental results of 8 widely studied SBFL formulae~\cite{zhang2017boosting} on the \toolsActual{} repair tools in terms of \dr{}. 
From the result, we have the following findings. First, we observe that \tech{} can reduce the patch validations by using all studied SBFL formulae on almost all repair tools. For example, studied SBFL formulae can achieve \dr{} ranging from 6.14\% (Kali-A - Ochiai2 / Tarantula) to 80.45\% (GenProg-A - Ochiai2) patch reductions for reaching the first plausible patch. Likewise, the studied SBFL formulae only degrades the performance of the \cardumen{} repair tool slightly (within 7.32\% to 13.41\%) with the same reason as discussed in Section~\ref{section:rq1}, i.e., \cardumen{} can generate plausible patches for only very few buggy versions, incurring unstable results.
Second, for different SBFL formulae, the overall patch reduction on all the studied \toolsActual{} tools is rather stable. 
For example, the formula with the best overall improvement is SBI (36.63\%) and the formula with the worst performance is Op2 (31.30\%), i.e., the difference of all studied formulae is only 5 percentage points \textbf{(\pp{}}).
Third, for each repair tool, the range of reductions among different formulae also seems small.
For example, for \jgenprog{}, the improvement for all formulae are the same (7.81\%).
For \kali, the best formula is SBI (6.36\%) while the worst is Ochiai / Tarantula (6.14\%). Such results demonstrate the effectiveness and stability of \tech{} when using different SBFL formulae. 

\begin{finding}{}
    All formulae under consideration achieve stable performance for \tech{}, boosting the overall validation results by at least 31.3\%. Meanwhile, the \textbf{SBI} formula performs the best for \tech{}, boosting validation by 36.63\%.
\end{finding}

%% file: paper-sections/research-questions/rq2/table.tex
\begin{table*}[tbh]
    \scalebox{1.0}{
        \centering
        \begin{tabular}{|c|c|c|c|c|c|c|c|c|c|c|c|c|}
            \hline
            \rowcolor{tablegray}  & Tarantula & Ochiai & Ochiai2 & Op2 & SBI & Jaccard & Kulczynski & Dstar2 \\
            \hline
            \hline
             \arja & 46.23\% & 40.05\% & 38.66\% & 33.57\% & 53.88\% & 39.91\% & 39.91\% & 40.60\% \\
            \rowcolor{tablegray} \avatar & 52.16\% & 54.80\% & 53.62\% & 51.74\% & 52.16\% & 55.22\% & 55.22\% & 53.55\% \\
             \cardumen & -7.32\% & -7.32\% & -7.32\% & -7.32\% & -13.41\% & -7.32\% & -7.32\% & -7.32\% \\
            \rowcolor{tablegray} \fixminer & 43.45\% & 43.55\% & 43.24\% & 46.46\% & 45.49\% & 43.45\% & 43.45\% & 43.65\% \\
             \genprog & 60.09\% & 78.81\% & 80.45\% & 78.21\% & 73.35\% & 78.81\% & 78.81\% & 78.91\% \\
            \rowcolor{tablegray} \jgenprog & 7.81\% & 7.81\% & 7.81\% & 7.81\% & 7.81\% & 7.81\% & 7.81\% & 7.81\% \\
             \jkali & 37.18\% & 37.18\% & 37.18\% & 37.20\% & 37.20\% & 37.18\% & 37.18\% & 37.18\% \\
            \rowcolor{tablegray} \jmutrepair & 20.26\% & 20.26\% & 20.26\% & 20.26\% & 20.26\% & 20.26\% & 20.26\% & 20.26\% \\
             \kali & 6.14\% & 6.36\% & 6.14\% & 6.36\% & 6.36\% & 6.36\% & 6.36\% & 6.36\% \\
            \rowcolor{tablegray} \kpar & 17.61\% & 17.56\% & 17.80\% & 14.20\% & 16.57\% & 17.56\% & 17.56\% & 17.56\% \\
             \rsrepair & 23.50\% & 21.84\% & 24.06\% & 21.32\% & 23.26\% & 23.21\% & 23.21\% & 21.84\% \\
            \rowcolor{tablegray} \tbar & 9.92\% & 10.93\% & 9.89\% & 6.52\% & 8.80\% & 10.01\% & 10.01\% & 11.16\% \\
 
            \hline
            \hline
             \textbf{Overall} & \textbf{33.55\%} & \textbf{35.54\%} & \textbf{35.54\%} & \textbf{31.30\%} & \textbf{36.63\%} & \textbf{35.54\%} & \textbf{35.54\%} & \textbf{35.61\%} \\
            \hline
        \end{tabular}
        }
        \caption{Impact of formula on \Comment{Displacement Ratio}patch reduction}
        \label{table:rq2}
\end{table*}

%% file: paper-sections/research-questions/rq3/content.tex
\subsection{RQ3: Impact of Granularity}
\label{section:rq3}

\input{paper-sections/research-questions/rq3/table}
\input{paper-sections/research-questions/rq3/tableNumQuality}
\input{paper-sections/research-questions/rq3/timings}


We now have studied \tech{} with patch-modification matrices at the default method level. 
In this section, we investigate the effectiveness of \tech{} on modification matrices with different granularities, i.e., prioritizing patches based on the modified code elements at different granularities (i.e., package, class, method, and statement levels).  
Table \ref{table:rq3}\Comment{ and Figure \ref{figure:rq3}} shows the \dr{} of \tech{} with modification matrices at the four granularities on the \toolsActual{} repair tools. 
From the table, we can observe that \tech{} with all the four granularities can reduce patch executions for most studied repair tools. More specifically, for the package and class levels, \tech{} can reduce patch executions for all repair tools, e.g., the patch executions for \genprog{} can be reduced by 35.08\% and 43.42\% at the package and class levels respectively. Interestingly, \tech{} tends to become less stable for finer granularities, e.g., \tech{} degrades the performance of one repair tool (\cardumen{}) at the method level and degrades the performance of two repair tools (\jgenprog{} and \kali{}) at the statement level.\checkLater{ The reason is that \lingming{put reason}}
We also observe that the overall patch reductions tend to be larger for finer granularities. For example, the overall reduction ratios for the package, class, method, and statement levels are 14.45\%, 17.84\%, 35.54\%, and 38.86\%, respectively. Also, \tech{} performs better at the method level than at package and class levels for 9/12 repair tools, while \tech{} at the statement level performs better than all other levels for 7/12 repair tools. The reason is that patch modification at finer granularities can capture more precise similarity information among different patches.


We also demonstrate the execution cost of \tech{}, shown in Table ~\ref{table:rq2-timing} for the four granularities. 
From the table, we can find that running \tech{} on all four granularities is extremely fast (e.g., less than 1 second on average for each buggy version), showing the practical usage of \tech{} within real-world tools / systems.
An interesting finding is that finer granularities do not cost obviously more time, e.g., the statement-level granularity cost even less time than other granularities. \checkLater{\sam{note that the implementation does quit early after observing the 1st plausible patch, may need statement of this}\lingming{fine without it now, leave for later}} The possible reasons are (1) our implementations have been highly optimized (shown in Section~\ref{section:alg}), and (2) \tech{} with finer granularities often prioritize plausible patches earlier and thus terminates with fewer iterations.

\begin{finding}{}
    \tech{} can substantially reduce patch executions for most studied repair tools with modification matrices at all four granularities; \tech{} with modification matrices at finer granularities tends to achieve higher overall reductions but also have larger variances. The \tech{} overhead is negligible at all four granularities. 
\end{finding}

%% file: paper-sections/research-questions/rq3/table.tex
\begin{table}[htb]
    \scalebox{1.0}{
        \centering
        \begin{tabular}{|c|c|c|c|c|}
            \hline
            \rowcolor{tablegray} & Package & Class & Method & Statement \\
            \hline
            \hline
            
             \arja & 15.09\% & 26.19\% & 40.05\% & 36.28\% \\
            \rowcolor{tablegray} \avatar & 0.56\% & 10.92\% & 54.80\% & 72.39\% \\
             \cardumen & 6.10\% & 1.22\% & -7.32\% & 43.90\% \\
            \rowcolor{tablegray} \fixminer & 2.50\% & 3.37\% & 43.55\% & 77.15\% \\
             \genprog & 35.08\% & 43.42\% & 78.81\% & 19.39\% \\
            \rowcolor{tablegray} \jgenprog & 31.25\% & 17.19\% & 7.81\% & -1.56\% \\
             \jkali & 24.36\% & 24.36\% & 37.18\% & 21.79\% \\
            \rowcolor{tablegray} \jmutrepair & 16.01\% & 16.01\% & 20.26\% & 28.76\% \\
            \kali & 27.19\% & 32.46\% & 6.36\% & -20.39\% \\
            \rowcolor{tablegray}  \kpar & 10.02\% & 8.63\% & 17.56\% & 48.36\% \\
            \rsrepair & 21.41\% & 18.66\% & 21.84\% & 31.93\% \\
            \rowcolor{tablegray}  \tbar & 4.99\% & 3.34\% & 10.93\% & 26.26\% \\
            
            \hline
            \hline
            \textbf{Overall} & \textbf{14.45\%} & \textbf{17.84\%} & \textbf{35.54\%} & \textbf{38.86\%} \\
            \hline
        \end{tabular}
    }
        \caption{Impact of granularity on \dr{}}
        \label{table:rq3}
\end{table}

%% file: paper-sections/research-questions/rq3/tableNumQuality.tex
\checkLater{
\begin{table}[htb]
    \scalebox{1.0}{
        \centering
        \begin{tabular}{|c|c|c|c|}

            \hline
            
            \rowcolor{tablegray} & \multicolumn{2}{c|}{\textbf{Number of Patches}} & \\
            
            \hhline{
            |
            >{\arrayrulecolor{tablegray}}-
            >{\arrayrulecolor{black}}|--|
            >{\arrayrulecolor{tablegray}}-
            >{\arrayrulecolor{black}}|
            }
            
        \rowcolor{tablegray} & Partial-matrix & Full-matrix & \multirow{-2}{*}{Percent Increase} \\
        \hline
        \hline
        
        \arja{} & 2567 & 4587 & 44.04\% \\
        \rowcolor{tablegray}\avatar{} & 167 & 200 & 16.50\% \\
        \cardumen{} & 39 & 98 & 60.20\% \\
        \rowcolor{tablegray}\fixminer{} & 51 & 51 & 0.00\% \\
        \genprog{} & 5177 & 6601 & 21.57\% \\
        \rowcolor{tablegray}\jgenprog & 7 & 44 & 84.09\% \\
        \jkali & 15 & 23 & 34.78\% \\
        \rowcolor{tablegray}\jmutrepair{} & 75 & 76 & 1.32\% \\
        \kali{} & 93 & 131 & 29.01\% \\
        \rowcolor{tablegray}\kpar{} & 503 & 532 & 5.45\% \\
        \rsrepair{} & 906 & 2619 & 65.41\% \\
        \rowcolor{tablegray}\tbar{} & 433 & 484 & 10.54\% \\
        
        \hline
        \hline
        \textbf{Overall} & \textbf{10033} & \textbf{15446} & \textbf{35.04\%} \\
        \hline

        \end{tabular}
    }
        \caption{Impact of Patch-Modification-Matrix on High Quality Patches}
        \label{table:rq3Num}
\end{table}}

%% file: paper-sections/research-questions/rq3/timings.tex
\begin{table}[tbh]
    \scalebox{1.0}{
        \centering
        \begin{tabular}{|c|c|c|c|c|}
            \hline
            \rowcolor{tablegray} & \multicolumn{4}{c|}{\textbf{Time (seconds)}} \\
            
            \hhline{
            |
            >{\arrayrulecolor{tablegray}}-
            >{\arrayrulecolor{black}}|----|
            }
            
            \rowcolor{tablegray} & Package & Class & Method & Statement \\
            \hline
            \hline
            
             \arja & 1.371 & 1.336 & 1.356 & 1.333 \\
            \rowcolor{tablegray} \avatar & 0.522 & 0.508 & 0.524 & 0.507 \\
             \cardumen & 0.291 & 0.286 & 0.285 & 0.280 \\
            \rowcolor{tablegray} \fixminer & 0.282 & 0.276 & 0.283 & 0.279 \\
             \genprog & 0.987 & 0.947 & 0.986 & 0.960 \\
            \rowcolor{tablegray} \jgenprog & 0.303 & 0.296 & 0.298 & 0.294 \\
             \jkali & 0.299 & 0.290 & 0.287 & 0.284 \\
            \rowcolor{tablegray} \jmutrepair & 0.299 & 0.287 & 0.288 & 0.282 \\
             \kali & 0.359 & 0.347 & 0.354 & 0.349 \\
            \rowcolor{tablegray} \kpar & 0.447 & 0.437 & 0.449 & 0.434 \\
             \rsrepair & 0.629 & 0.604 & 0.620 & 0.604 \\
            \rowcolor{tablegray} \tbar & 0.519 & 0.501 & 0.521 & 0.502 \\
            
            \hline
            \hline
             \textbf{Average} & \textbf{0.526} & \textbf{0.510} & \textbf{0.521} & \textbf{0.509} \\
            \hline
        \end{tabular}
        }
        \caption{Overhead of \tech{}}
        \label{table:rq2-timing}
\end{table}

%% file: paper-sections/research-questions/rq4/content.tex
\subsection{RQ4: Impact of Full Validation Matrix}
\label{section:rq4}

\input{paper-sections/research-questions/rq4/table}


We now have studied \tech{} with the default partial patch-validation matrices.
In this section, we further investigate the performance of \tech{} with full patch-validation matrices. 
Table~\ref{table:rq4} shows the comparison between \tech{} with partial and full patch-validation matrices. 
From this table, we see that \tech{} can achieve significant reductions on both full and partial patch-validation matrices, indicating the general applicability of \tech. More specifically, the overall reductions by \tech{} with the default partial matrices is even larger than \tech{} with full matrices (35.54\% vs 20.37\%). 
For example, \tech{} substantially degrades the performance of some repair tools when using full matrices, e.g., the \dr{} on \arja{} changes from 40.05\% (with partial matrices) into -16.27\% (with full matrices). 

\checkLater{One possible reason for why \tech{} performs worse with full matrices is because a significant portion of low-quality patches with partial matrices are considered high-quality patches with full matrices e.g. from 25997 patches to 4587 patches for \arja{}, shown in Table \ref{table:rq3Num}. In fact, the only subjects affected by partial/full matrices are those with multiple originally failing tests. Only 3 subjects fit this criteria for \arja{}, Chart-25, Math-68, and \highlight{Math-95 (point of discussion only 1 ori. failing test according to d4j}. The ranks of the earliest plausible patch for these subjects change from 4 - >5, 278 -> 1383, and 4 -> 958 respectively. Thus we see that the performance for most subjects remain consistent, but subjects with multiple originally failing tests may suffer huge degradation.
Another interesting finding is that for some repair tools, the performance of \tech{} is same on partial and full matrices (e.g., 54.8\% for \avatar{} and 37.18\% for \kali{}). We find this to be due very little variance in the number of high-quality patches produced between the different patch matrices (e.g. 5.45\% increases for \kpar{}). In fact, according to Table \ref{table:rq3Num}, we observe a trend where tools that with lowest percent increases in high-quality patches remain the most consistent between patch-modification-matrices!
}

\begin{finding}{}
   \tech{} can effectively reduce patch executions on both partial and full patch-validation matrices. \tech{} tends to perform even better with partial matrices than with full matrices.
\end{finding}

%% file: paper-sections/research-questions/rq4/table.tex
\begin{table}[htb]
    \scalebox{1.0}{
        \centering
        \begin{tabular}{|c|c|c|}
            \hline
            \rowcolor{tablegray}  & Partial-matrix & Full-matrix \\
            \hline
            \hline

             \arja & 40.05\% & -16.27\% \\
            \rowcolor{tablegray} \avatar & 54.80\% & 54.80\% \\
             \cardumen & -7.32\% & -10.98\% \\
            \rowcolor{tablegray} \fixminer & 43.55\% & 43.55\% \\
             \genprog & 78.81\% & 50.05\% \\
            \rowcolor{tablegray} \jgenprog & 7.81\% & -3.13\% \\
             \jkali & 37.18\% & 26.92\% \\
            \rowcolor{tablegray} \jmutrepair & 20.26\% & 7.19\% \\
             \kali & 6.36\% & 10.09\% \\
            \rowcolor{tablegray} \kpar & 17.56\% & 17.56\% \\
             \rsrepair & 21.84\% & 19.19\% \\
            \rowcolor{tablegray} \tbar & 10.93\% & 10.93\% \\

            \hline
            \hline
            \textbf{Overall} & \textbf{35.54\%} & \textbf{20.37\%} \\
            \hline
        \end{tabular}
    }
    
    \caption{Impact of validation matrix on \dr{}}
    \label{table:rq4}
\end{table}

%% file: paper-sections/research-questions/rq6/content.tex
\subsection{RQ5: Impact of Additional Patch Similarity Information}

In this research question we examine the impact of utilizing additional information for computing patch similarity. We execute the default configuration of \tech{} and compute patch similarity using both (1) the set of modified elements and (2) the applied patch templates or patterns. We achieve this by looking specifically only at tools which apply predefined repair patterns, i.e., those tools categorized as \textit{template-based}. Note that we collect the applied fix patterns as directly reported by each tool.
The studied template-based tools are \tbar{}, \kpar{}, \avatar{}, and \fixminer{} with 18, 13, 19, and 23 distinct repair patterns, respectively.

Table \ref{table:rq6} shows the result of this configuration evaluated against four patch-modification matrix granularities.
In this table, each repair tool name followed by symbol "++" represents the results for our new \tech{} variant with additional patch pattern information, i.e., \techPat{}.
From the table, we can clearly find that \techPat{} with additional patch pattern information performs much better than \tech{} for all studied repair tools on all four different granularities.
On \avatar{} for instance, \techPat{} can achieve a reduction of 63.00\% at the package granularity while \tech{} only achieves a reduction of 0.56\%; similarly for \tbar{}, \techPat{} can achieve a reduction of 36.65\% at the package granularity while \tech{} only achieves a reduction of 4.99\%.
Another interesting finding is that the improvement achieved by \techPat{} tends to be larger when using coarse-grained granularities than when using fine-grained granularities. The reason is that the original \tech{} computes very coarse-grained patch similarity information when using coarse-grained granularities, leaving more room for \techPat{} with additional fixing pattern information to further boost the patch reduction. \checkLater{\lingming{what is the small purple ''3'' right after this paragraph? remove it}}

\checkLater{\footnote{\highlight{We currently exclude \fixminer{} because ...}}}

\input{paper-sections/research-questions/rq6/table}

\begin{finding}
    Fixing pattern information further boosts \tech{}'s performance on all the studied repair tools (e.g., by up to 62.44 pp). Fixing pattern information tends to work better with coarser-grained granularities.
\end{finding}

%% file: paper-sections/research-questions/rq6/table.tex
\begin{table}[tbh]
    \scalebox{0.95}{
        \centering
        \begin{tabular}{|c|c|c|c|c|}
            \hline
            \rowcolor{tablegray}  & Package & Class & Method & Statement \\
            \hline
            \hline
            
             \avatar & 0.56\% & 10.92\% & 54.80\% & 72.39\% \\
            \rowcolor{tablegray} \avatar{}++ & 63.00\% & 63.63\% & 80.53\% & 78.09\% \\
            \hline
            
            
             \kpar & 10.02\% & 8.63\% & 17.56\% & 48.36\% \\
            \rowcolor{tablegray} \kpar{}++ & 32.91\% & 32.75\% & 33.33\% & 50.39\% \\
            \hline
            
             \tbar & 4.99\% & 3.34\% & 10.93\% & 26.26\% \\
            \rowcolor{tablegray} \tbar{}++ & 36.65\% & 37.25\% & 40.65\% & 34.11\% \\
            \hline
        \end{tabular}
    }
    \caption{\tech{} with additional patch information}
    \label{table:rq6}
\end{table}

%% file: paper-sections/research-questions/rq5/content.tex
\subsection{RQ6: Boosting \tech with Other \apr Tools}
\label{section:rq5}

\input{paper-sections/research-questions/rq5/table}

Table \ref{table:rq5} shows the impact of historical information for our default configuration. Note, we filter out all known correct patches from this historical information, since developers will stop the repair process after finding any correct patch, thus alleviating the need for \tech{}. Compared to Table \ref{table:rq1}, we observe improvements across most repair tools, up to 65.35 \pp{} (\kali{}). Most notably, the only tool with originally negative performance (\cardumen{}) now has 30.49\% reduction, i.e., over 37 \pp{} improvement. \Comment{\sam{current text-based on outdated information, perhaps focus on tools with poor performance (cardumen, jgenprog, and kali-a) instead of those with negative performance)}}

Likewise, the only two instances of degradation comes from \genprog{} and \tbar{}, e.g., \genprog{} degrades from 78.81\% to 63.50\%, while \tbar{} degrades from 10.93\% to -1.59\%. Closer inspection reveals that one reason for such degradation mainly comes from specific projects rather than consistent degradation over all projects. For example, for \tbar{}, the performance for Chart drops by over 57 \pp{}\Comment{ and 16.11 \pp{}\highlight{check numbers on all claims}, respectively, causing the overall degradation}. Meanwhile, for \tbar{}, \tech{} can still be massively effective on other projects, e.g. the average rank on the Time subject over 2 versions decreases from 64.5 to 13.5, a performance increase from -174.47\% to 44.68\% \checkLater{\lingming{different with the table, which is 44.68\%, please double check all nums globally}}.

\Comment{The average new rank for these subjects are 73.89 from 67.89 over 9 subjects and 24.33 from 22.58 over 12 subjects for Chart and Closure respectively. \lingming{we should explain the reason here}. \tbar{}'s performance degrades\lingming{unfinished sentence?} \sam{Still, \tech{} can still be massively effective, e.g. the average rank on Time decreasing from 64.5 to 13.5 over 2 subjects, a performance increase from -174.47\% to 43.75\%.}}


\begin{finding}{}
    Supplementing \tech{} with patch-execution information from other repair tools can further boost the overall performance of \tech{}. This extra historical information, compared to \tech{}'s default configuration, can further boost the results by 21.43 \pp{} on average and up to 65.35 \pp{}.
\end{finding}

%% file: paper-sections/research-questions/rq5/table.tex
\begin{table*}[tbh]
    \scalebox{0.85}{
        \centering
        \begin{tabular}{|c|c|c|c|c|c|c|c|c|c|c|c|c|c|}
        \hline
        \rowcolor{tablegray}  & \arja{} & \avatar{} & \cardumen{} & \fixminer{} & \genprog{} & \jgenprog{} & \jkali{} & \jmutrepair{} & \kali{} & \kpar{} & \rsrepair{} & \tbar{} \\
        \hline
        \hline
        
         Chart & 87.06\% & 28.20\% & 24.24\% & 57.84\% & 85.08\% & 90.00\% & 0.00\% & 27.27\% & 55.24\% & 45.15\% & 84.47\% & -8.67\% \\
        \rowcolor{tablegray} Closure & --- & 67.36\% & --- & --- & --- & --- & --- & --- & --- & 63.49\% & --- & -7.38\% \\
         Lang & 52.31\% & 53.76\% & 32.00\% & 42.55\% & -538.30\% & -200.00\% & 48.15\% & 24.10\% & --- & 1.94\% & 16.25\% & -63.06\% \\
        \rowcolor{tablegray} Math & 80.03\% & --- & 37.50\% & --- & 71.77\% & 70.00\% & 47.62\% & 12.50\% & 76.64\% & 27.43\% & 76.30\% & 38.53\% \\
         Mockito & --- & 25.00\% & --- & --- & --- & --- & --- & --- & --- & 55.49\% & --- & 17.95\% \\
        \rowcolor{tablegray} Time & --- & 94.89\% & --- & 13.95\% & --- & --- & --- & --- & --- & 68.27\% & --- & 44.68\% \\
         
        \hline
        \hline
         \textbf{Overall} & \textbf{83.68\%} & \textbf{62.10\%} & \textbf{30.49\%} & \textbf{47.73\%} & \textbf{63.50\%} & \textbf{62.50\%} & \textbf{42.31\%} & \textbf{20.92\%} & \textbf{71.71\%} & \textbf{38.29\%} & \textbf{67.31\%} & \textbf{-1.59\%} \\
         \hline
        \end{tabular}
    }
    \caption{\tech{} historical results}
    \label{table:rq5}
\end{table*}

%% file: paper-sections/07_conclusion.tex
\section{Conclusion}
\label{section:conclusion}

We have proposed the first self-boosted \apr technique, \tech, which leverages the execution information of validated patches during \apr{} to prioritize the remaining patches on-the-fly for faster \apr. In particular, for each remaining patch, \tech{} infer its similarity with the validated high/low-quality patches and promotes/degrades it accordingly to speed up plausible/correct patch generation. We conducted an extensive study to evaluate \tech{} on \toolsActual{} state-of-the-art \apr{} systems and the widely used Defects4J benchmark suite. The experimental results demonstrate that the default \tech{} can overall reduce 35.52\% patch validations with negligible overhead (e.g., 0.5s on average). We further studied the impact of different configurations on \tech, and observed that (1) \tech{} has stable performance when using different SBFL formulae for patch prioritization, (2) \tech{} can effectively reduce patch executions for most studied repair tools with patch-modification matrices at all four granularities (e.g., package, class, method and statement levels) and \tech{} with finer-grained modification matrices tends to achieve better overall improvement, (3) \tech{} works for both partial/full patch-validation matrices and tends to perform better with partial matrices, (4) \tech{} can be further improved by utilizing additional patch pattern information to compute patch similarity, and (5) \tech{} can perform even better by leveraging patch-execution information from other repair tools.
